\documentclass[aps,pra,reprint,10pt,a4paper,nofootinbib]{revtex4-2}
\usepackage[utf8]{inputenc}
\usepackage[sc,osf]{mathpazo}
\usepackage{amsmath}
\usepackage[T1]{fontenc}
\usepackage{latexsym}
\usepackage{amssymb}
\usepackage{tabularx}
\usepackage[colorlinks=true,citecolor=blue,urlcolor=blue]{hyperref}
\usepackage{color}
\usepackage{graphics,epstopdf}
\usepackage{subcaption}
\usepackage{soul}
\usepackage{graphicx}
\usepackage{capt-of}
\usepackage{lipsum}
\usepackage{adjustbox}
\usepackage[normalem]{ulem}
\usepackage[table,xcdraw]{xcolor}
\usepackage{braket}
\usepackage{physics}
\usepackage{ragged2e}
\usepackage{mathtools}

\usepackage{comment}
\usepackage{ragged2e}
\usepackage{multirow}
\usepackage{braket}
\usepackage[table]{xcolor}
\usepackage{array}
\usepackage[font=small,labelfont=bf,justification=justified,singlelinecheck=false]{caption}
\definecolor{indiagreen}{rgb}{0.07, 0.53, 0.03}
\definecolor{teal}{rgb}{0.0, 0.53, 0.53}

\begin{document}

\title{Noise adaptive two-way secure  deterministic quantum key distribution}

\author{Abinash Kar\(^1\), Ayan Patra\(^{2,3}\), Aditi Sen(De)\(^{2,3}\), and Tamoghna Das\(^1\)}

    \affiliation{\(^1\)Department of Physics, Indian Institute of Technology Kharagpur, Kharagpur 721302, India}
	\affiliation{\(^2\)Harish-Chandra Research Institute,  Chhatnag Road, Jhunsi, Prayagraj 211 019, India}
    \affiliation{\(^3\)Homi Bhabha National Institute, Training School Complex, Anushakti Nagar, Mumbai 400 094, India}
	
\begin{abstract}

We introduce \emph{noise-adaptive quantum key distribution (QKD)} protocols, in which the honest parties optimize the encoding (state preparation) and decoding (measurement basis) operations according to the noise models affecting the honest subsystems induced by an eavesdropper. This extends conventional QKD schemes that employ fixed encoding and decoding strategies independent of the noise characteristics of the communication channel. We investigate three representative protocols: entanglement-based secure dense coding (SDC), the entanglement-free Lucamarini and Mancini (LM05), and a two-way prepare-and-measure Bennett Brassard (BB84) protocols. Using entropic uncertainty relations, we derive the corresponding secret key rates for both adaptive and conventional non-adaptive scenarios under collective attacks. For independent but identical noise acting on the forward and backward transmission channels, as well as for correlated and non-Markovian environments, we identify classes of channels for which adaptive schemes yield enhanced secret key rates for the considered protocols. In contrast,  we also determine Pauli channels, including depolarizing and bit flip channels, for which adaptive strategies provide no benefit. We further show that these optimal sets are generally non-unique and can differ substantially from the unitaries that maximize dense-coding capacity in the absence of security constraints. Our results establish noise-adaptive encoding and decoding as a powerful framework for improving secure communication over realistic noisy quantum channels.

\end{abstract}
\maketitle

    

\section{Introduction}


The rapid development of quantum technologies, mainly in the direction of quantum algorithms,~\cite{Nielsen_2010_book, Deutsch1985, Shor_Algorithm, Grover1996, Preskill2018quantumcomputingin} eventually undermines the existing classical  cryptographic schemes~\cite{RSA1978}, based on computational complexity~\cite{Nielsen_2010_book}, leading to the search for fundamentally secure alternatives~\cite{BB84, Bennett_TCS_2014_BB84, Ekert_PRL_1991_Ekert-protocol, Mayers98, Lo_Science_1999_unconditional-security,Shor_PRL_2000_device-dependent-BB84, Gisin_RMP_2002_cryptography-review,
Scarani_RMP_2009_cryptography-review}. 
Quantum key distribution (QKD) addresses this challenge by exploiting intrinsic quantum features,  such as superposition and nonclassical correlations~\cite{Bell, CHSH, Bell-nonlocality}, particularly entanglement~\cite{Horodecki_RMP_2009_entanglement-review}, to enable information-theoretically secure \cite{Csiszar_IEEE_1978_Csiszr-Korner-individual-key-rate, Maurer1993} key generation between distant parties, even against adversaries with unbounded computational power.  Such security can be achieved in both device-dependent~\cite{BB84, Shor_PRL_2000_device-dependent-BB84, Mayers98,Lo_Science_1999_unconditional-security, Gisin_RMP_2002_cryptography-review} and device-independent scenarios~\cite{Ekert_PRL_1991_Ekert-protocol, Acin_PRL_2007_device-independent, Masanes_Nature-Comm_2011_device-independent-security}, provided the adversary is constrained only by the laws of quantum mechanics.
The earliest QKD protocol, BB84~\cite{BB84,  Bennett_TCS_2014_BB84}, along with its modified versions~\cite{Bennett1992, Bruss_PRL_1998_six-state, Bechmann_PRA_1999_six-state}, demonstrate that encoding information in non-orthogonal bases allows the legitimate users, Alice and Bob, to detect  eavesdropping attempts and distribute secure keys even in the presence of collective attacks~\cite{Shor_PRL_2000_device-dependent-BB84}. Shortly thereafter, an entanglement-based protocol (Ekert91) \cite{Ekert_PRL_1991_Ekert-protocol} is introduced where security can be certified through the violation of Bell inequalities. These protocols laid the foundations of both prepare-and-measure and entanglement-assisted QKD, which have since been strengthened through rigorous unconditional security proofs, including those by Mayers \cite{Mayers98}, Lo–Chau \cite{Lo_Science_1999_unconditional-security}, and Shor–Preskill \cite{Shor_PRL_2000_device-dependent-BB84}.

Alongside these theoretical advances, remarkable experimental progress enables long-distance QKD implementations with continuous improvements in robustness and security~\cite{Stucki_2002,Liao_Nature_2017_satellite-QKD-1200km, Liao_PRL_2018_satellite-QKD-7600km, Fang_NP_2020_QKD-500km, Chen_prl_2020_Laser-QKD-500km}. 
For example, decoy-state QKD mitigates photon-number-splitting attacks in weak coherent pulse systems~\cite{Decoy-state-PhysRevLett.91.057901, Schmitt-Manderbach_PRL_2007_decoy-state-QKD-polarisation-144km, Peng_PRL_2007_decoy-state-QKD-polarization}, while measurement-device-independent QKD (MDI-QKD) eliminates all detector-side vulnerabilities by outsourcing measurements to an untrusted relay~\cite{MDI-PhysRevLett.108.130503, Yin_PRL_2016_404km-QKD, Pirandola_NP_2015_CV-MDI-QKD-experiment}. On the other hand, device-independent QKD (DI-QKD) derives security solely from observed violations of Bell inequality  without requiring detailed knowledge of the internal functioning of the devices~\cite{Acin_PRL_2007_device-independent, Masanes_Nature-Comm_2011_device-independent-security, Pironio_PRX_2013_device-independent-security, Vazirani_PRL_2014_fully-device-independent} (see also Ref.~\cite{Kent13,NSDI-Acin-PhysRevLett.97.120405,Masanes2014, Winczewski-PhysRevA.106.052612} for DI security against even more powerful adversaries).
 
 At the same time, deterministic two-way QKD protocols, such as the ping-pong protocol~\cite{Bostrom_PRL_2002_ping-pong}, Lucamarini and Mancini (LM05)~\cite{Lucamracini_PRL_2005_LM05}, and their higher-dimensional~\cite{Muhuri_PLA_2026} or entanglement-assisted variants~\cite{Beaudry_PRA_2013_two-way-QKD, Dimensional_advantage},  explore the bidirectional use of quantum channels for secure communication. In secure dense-coding (SDC)-based schemes~\cite{Bostrom_PRL_2002_ping-pong,Beaudry_PRA_2013_two-way-QKD,Dimensional_advantage}, the honest parties can effectively double the shared key rate by exploiting one shared ebit of entanglement together with the two-way use of the quantum channel. This enables deterministic key generation without basis sifting and can enhance the detectability of eavesdropping because the information-carrying qubit traverses the channel twice. A related idea is realized in the entanglement-free LM05 protocol~\cite{Lucamracini_PRL_2005_LM05,Lu_PRA_2011_LM05-secure}, where one party prepares a qubit in one of the two conjugate bases and the other encodes information through unitary operations before returning the qubit. Contrary to the BB84, where the raw key is generated probabilistically, both LM05 and SDC allow deterministic encoding and decoding of information on traveling qubits.

Despite these developments, a central limitation of all these QKD protocols is that the encoding operations, basis choices, and decoding measurements are typically fixed in advance, independent of the underlying channel noise or eavesdropping strategy. This rigidity contrasts sharply with results from quantum dense coding in noisy environments~\cite{Shadman2010, Shadman2011, Shadman2012,Das_PRA_2014_noise-inverts-dense-coding,Das_distributed-PhysRevA.92.052330}, where the optimal encoding strategy depends explicitly on the characteristics of the noise. 
Since collective eavesdropping attacks in QKD can always be modeled as completely positive trace-preserving (CPTP) maps acting on the transmitted subsystem~\cite{Nielsen_2010_book}, fixed encoding strategies are, in general, not expected to maximize secure key rates.

This observation motivates to introduce the concept of \emph{noise-adaptive} QKD (NAQKD), in which the honest parties can dynamically optimize the encoding unitaries and decoding measurements according to the estimated noise acting on the transmitted quantum states. Since both environmental noise and adversarial attacks effectively manifest as CPTP maps on the transmitted subsystem, adapting the encoding strategy can enhance both information transmission and security. 
In this work, we analyze three representative QKD protocols, entanglement based secure dense coding (SDC) \cite{Beaudry_PRA_2013_two-way-QKD, Dimensional_advantage}, entanglement free LM05 \cite{Lucamracini_PRL_2005_LM05, Muhuri_PLA_2026}, and a two-way version of prepare-and-measure BB84 protocol, and derive the corresponding secret key rates for both the adaptive and the conventional strategies under collective eavesdropping attack, with the help of entropic uncertainty relation \cite{Berta_NP_2010_uncertainty-principle-memory}. When independent but identical noise affects both the forward and backward transmission channels, we identify certain classes of unital channels exhibiting no-gain behavior, while significant enhancement can be observed for phase flip,  bit-phase flip 
and non-unital noise models, such as the amplitude damping channel. In particular, bit-phase flip noise shows an advantage of the adaptive method for all three protocols; phase flip noise shows advantage for SDC only, and amplitude damping finds improved key rate for LM05 and two-way BB84 protocols.

Going beyond independent Markovian noise models, we illustrate that partially correlated bit-phase flip and phase flip channels can still provide advantages for adaptive schemes, whereas fully correlated noise typically leads to a no-gain regime. Similar enhancements in secret key rates are also observed in the presence of non-Markovian environments. Throughout our analysis, we further identify the classes of encoding unitaries that yield higher key rates in adaptive protocols compared to their conventional non-adaptive counterparts. Our results reveal that the optimal adaptive unitaries are generally non-unique; nevertheless, Clifford unitaries are sufficient in most scenarios to achieve the observed enhancement. Interestingly, we also find that the unitaries optimizing noisy dense coding do not necessarily coincide with those that maximize the secure key rate in secure dense-coding protocols.



The paper is organized as follows: In
Sec. \ref{Sec:Generic-noise-adaptive-protocol}, we introduce the noise adaptive QKD protocol for a generic two-way protocol, followed by the purified version of that protocol in Sec. \ref{sec:purified-generic}. Subsequently, in Sec. \ref{sec:generic-key-rate},  we derive the closed-form expression of the secure key rates of these generic protocols with the help of the entropic uncertainty relation. In the presence of various noise models, we explicitly calculate and compare the adaptive and non-adaptive key rates for three major QKD protocols in Sec. \ref{sec:results-noise-adaptive}, when independent and identical noise acts in the transmission channel.  In Sec. \ref{sec:correlated-channel}, we investigate the secret key rate when fully and partially correlated channels act on honest parties, while the advantages of the proposed protocols in the presence of non-Markovian channels are discussed in Sec. \ref{sec:nonMarkovian}. In Sec. \ref{sec:DCcapacity-Comp}, we compare optimal unitaries for the adaptive SDC protocol with the dense coding   capacity obtained without imposing security constraints. The concluding remarks are included in Sec. \ref{sec:conclusion}.

\section{Generic noise adaptive two-way deterministic quantum key distribution protocol}\label{Sec:Generic-noise-adaptive-protocol}

A two–way deterministic quantum key distribution protocol, $\mathcal{P}$, distributes secure quantum keys between the two distant honest parties, Alice and Bob, with the help of a pre-shared bipartite entangled state $\rho_{AB}$ in $\mathcal{H}^2_A \otimes \mathcal{H}^2_B $\footnote{The superscripts are used to specify the dimension of the Hilbert spaces.}. It is majorly composed of three steps: (1) the preparation of a bipartite signal state $\rho_{AA'}$, 
(2) an encoding unitary set $\{U^k\} \in \mathcal{U} \equiv SU(2)$, used to encode the raw key value $k$,
and (3) a decoding positive operator valued measure (POVM) $\{M^{l}\} \in \mathcal{M}$, to retrieve the encrypted raw key bit value. 
In this protocol, one of the two legitimate parties prepares an entangled two-qubit state, retains one qubit in a quantum memory, while sending the other one, for back-and-forth transmission, which involves the encoding of secret raw keys. On the other hand, for the decoding of the secure keys, a joint measurement is performed on both the qubits. 

While the qubit travels through a quantum channel, it inevitably interacts with the  environment,
thereby affecting the key transmission probabilities. Such noise can be due to decoherence or the consequence of an eavesdropping attack. In this article, we develop a novel key distribution protocol for two-way communication, in which one of the honest parties is allowed to choose a generalized unitary operation for encoding of the secret raw keys, while the other one performs a generalized measurement for decoding the keys. The choice of these generalized encoding and decoding operations is determined by the possible noise models influencing the quantum transmission channel. We refer to this framework as a \emph{noise-adaptive} (\emph{noise-resilient}) quantum key distribution protocol (NAQKD\footnote{In this article, all the QKD protocols considered are two-way.}), as it optimizes the 
encoding and decoding strategies according to the underlying noise model of the channel, which in turn outputs the maximum amount of secure key rate for a given class of noise.





For a given noise model affecting the forward and backward transmission channels involved in the two-way physical transmission of the qubit, the noise-adaptive secure two-way quantum key distribution protocol proceeds as follows:

    $(1)$ Alice, one of the legitimate parties, prepares the maximally entangled Bell state, $|\phi^+\rangle_{AA'} = \frac{1}{\sqrt{2}}(|00\rangle + |11\rangle)_{AA'}$, keeps one qubit ($A$) in her quantum memory, and sends the other one ($A'$) to Bob via a forward quantum transmission channel $\Lambda^f_{A' \rightarrow B}$, resulting in the shared state $\rho_{AB} =\Lambda^f_{A' \rightarrow B} (\rho_{AA'})$, where $\rho_{AA'} = |\phi^+\rangle \langle\phi^+|_{AA'}$. 
    
    $(2)$ Upon receiving the shared part, the other honest party, Bob, eventually performs either the key generation run or the control check (to detect the possible eavesdropping) run. The key generation runs that Bob performs with a probability $c \approx 1$, consist of applying local unitary operations $\{U^{xy}\}_{x,y = 0}^1$ 
    chosen uniformly at random probabilities, 
    on his part to encode the classical raw key pair $(x,y)$, where $x,y\in\{0,1\}$.
    Unlike the existing two-way protocols \cite{Beaudry_PRA_2013_two-way-QKD, Lu_PRA_2011_LM05-secure}, where the encoding $U^{xy}$ is restricted to a fixed set of Pauli operations, the introduced protocol here allows Bob to employ arbitrary unitary operators in $SU(2)$, thereby ensuring noise-adaptive optimization of the encoding strategy.
    In this step, Bob's unitary operation leads to a shared ensemble, $\{p^{xy} = \frac 14,\rho_{AB}^{xy}\}$, where, $\rho_{AB}^{xy}=(\mathbb{I}_A\,\otimes U^{xy}_B)\,\rho_{AB}\,(\mathbb{I}_A\,\otimes U^{\dagger\,xy}_B)$.
    
    $(3)$ Bob sends back the encoded qubit to Alice through another backward quantum transmission channel $\Lambda^b_{B\rightarrow A'}$, thereby resulting in each member state of the ensemble to $\{p^{xy},\rho_{AA'}^{xy}\}$. This step can be represented as $\Lambda^b_{B \rightarrow A'} (\rho_{AB}^{xy}) = \rho_{AA'}^{xy}$. 
    
    $(4)$ Alice again randomly chooses two operations, one is for the key generation run with a very high probability, $c \approx 1$, and the other one is for a test run with the remaining probability ($1 - c$). She performs a joint measurement $\mathcal{M}_{AA'} = \{M_{AA'}^{ij}\}$ on both the qubits to extract maximum possible information about Bob's encoded key pairs $(x,y)$. Note that the choice of this decoding measurement, $\mathcal{M}_{AA'}$, completely depends on the encoding operations performed by Bob and the character of the forward and backward transmission channels. 

In the absence of noise or eavesdropping in the forward transmission, the shared state between the legitimate parties remains the pure state $\ket{\phi^+}_{AB}$. Any deviation from this ideal scenario, manifested as noise in the channel, may be attributed to a potential eavesdropping attack. In this work, we establish the security of the proposed noise-adaptive key distribution protocol against collective attacks. Specifically, in the case of a collective attack, an adversary, Eve, is assumed to possess quantum systems correlated with $N$ independently and identically distributed copies of the bipartite states shared between the honest parties over $N$ rounds of the protocol. Further, she is allowed to perform an optimal joint measurement on all her subsystems, which may be deferred until after the completion of the one-way classical post-processing stage, thereby enabling her to exploit all publicly revealed information.

To ensure security in the worst-case scenario, we adopt a purification-based approach and grant the adversary access to all possible additional interfaces or auxiliary systems associated with the purification of the shared state $\rho_{AB}$. This corresponds to assigning the eavesdropper maximal operational power achievable within the realm of quantum mechanics.

To estimate the level of noise in the transmission channel, or equivalently, to bound the correlations of Eve's subsystem, the honest parties, Alice and Bob, perform, in addition to the key generation rounds (which occur with probability arbitrarily close to unity), randomly interspersed security check (test) rounds. Eve can attack the quantum transmission channel in two different ways: $(i)$ She may tamper with the traveling qubit before the encoding operation performed by Bob, thereby altering the shared entangled state; $(ii)$ She can attack the backward channel after Bob's encoding to extract information about the shared key bit string. 
Hence, to determine the presence of Eve, and to ensure security, the test run performed by the honest parties comprises the following steps:

    $(1)$ Bob performs a projective measurement of the spin observable $\hat{n}\cdot\vec{\sigma}$ on his qubit, corresponding to a measurement along an arbitrary direction $\hat{n}$. Conditioned on the measurement outcome, associated with the eigenvalues $\pm 1$, he prepares and sends the corresponding eigenstate of a (generally different) rotated spin observable $\hat{m}\cdot\vec{\sigma}$ to Alice through the backward quantum channel $\Lambda^{b}_{B\rightarrow A'}$.

    $(2)$ After receiving Bob's input, Alice measures her stored qubit in the same $\hat{n}\cdot\vec{\sigma}$-basis, like Bob, but measures the received qubit in a different $\hat{m'}\cdot\vec{\sigma}$-basis. The choice of the two different unit vectors $\hat{n}$ and $\hat{m'}$ is decided by the honest parties prior to the protocol, and it is completely based on the noise models acting on the transmission channels. 

After a sufficiently large number of quantum key distribution protocol rounds, involving both the key generation and the test runs performed by the honest parties randomly, any one of the honest parties starts a one-way classical post-processing protocol, which involves shifting of keys, classical error corrections, and privacy amplification. During this process, the honest parties estimate the possible lower bound on the secret key rate of their protocol, according to the Devetak-Winter bound~\cite{Devetak_PRSA_2005_key-rate}. If the estimated key rate turns out to be positive, they proceed further with the post-processing events; otherwise, they abort the protocol. 


In the following, we present the security analysis of the proposed noise-adaptive protocol and derive a lower bound on the achievable secret key rate. To this end, we first describe a purified version of the noise-adaptive two-way protocol and establish its equivalence with the above-stated protocol. 



\subsection{Purified NAQKD protocol}\label{sec:purified-generic}

Due to potential eavesdropping attack in the forward transmission channel, the honest parties end up by sharing mixed $\rho_{AB}  = \Lambda^f_{A' \rightarrow B}(\rho_{AA'})$, irrespective of $\rho_{AA'} = |\phi^+ \rangle \langle\phi^+|_{AA'}$, a maximally entangled Bell state. 
The encoding operation by Bob using an arbitrary set of unitaries $U^{xy}$, where $x,y \in \{0,1\}$, is needed to be optimized to maximize the key rate.
We will now prove that the unitary encoding $U^{xy}$ can be equivalently purified to a measurement on an auxiliary system. Suppose Bob introduces an auxiliary maximally entangled Bell state, $|\phi^+\rangle$, performing a complete set of joint von-Neumann measurements $F^{xy} \in \mathbb{C}^2 \otimes \mathbb{C}^2$, where $\sum_{xy} F^{xy} = \mathbb{I}$, on one signal qubit and another auxiliary qubit. Hence, this deterministic encoding, performed by Bob, can be equivalently described as measurement-induced entanglement swapping \cite{PhysRevLett.71.4287, PhysRevA.57.822}, which is mathematically expressed as  
\begin{widetext}
\begin{equation}\label{eq:purification-general}
    4 \times {}_{XX'}\langle \phi(xy)|\left(\rho_{AX} \otimes |\phi^{+}\rangle\langle \phi^{+}|_{X'B}\right)|\phi(xy)\rangle_{XX'} = (\mathbb{I}_{A}\otimes U_B^{xy} )\rho_{AB} (\mathbb{I}_{A}\otimes U_B^{xy\,\dagger} ) \equiv \rho_{AB}^{xy}
\end{equation}
\end{widetext}
where 
\begin{align}
    &|\phi^{+}\rangle_{X'B} = \frac{1}{\sqrt{2}}(|00\rangle + |11\rangle)_{X'B}= \frac{1}{\sqrt{2}} \sum_{p=0}^{1}|p,p\rangle_{X'B} \label{eq:aux}
\end{align}   
is the maximally entangled state in the composite $4$-dimensional ($\mathbb{C}^2 \otimes \mathbb{C}^2$) complex Hilbert space, and, 
\begin{align}
    &|\phi(xy)\rangle_{XX'} = \frac{1}{\sqrt{2}} \sum_{l,m=0}^{1} (U^{xy})^\dagger_{ml}|m,l\rangle_{XX'} \label{eq:Bellwhat}\,,
\end{align}
with $(U^{xy})_{ml} = \langle m|U^{xy}|l\rangle$, is the $m,l~$th matrix element of the unitary operator $U^{xy}$. The proof of Eq. \eqref{eq:purification-general} is given in Appendix \ref{app:generalized-purification} 
(cf. \cite{Bennett_PRL_1993_teleportation}).

It is worth mentioning the fact that the set of all possible measurements $\{|\phi(xy) \langle\phi(xy)|\}_{x,y = 0}^1$, forms a complete set of basis in the $\mathbb{C}^2 \otimes \mathbb{C}^2$ dimensional Hilbert space, i.e., $\sum_{x,y = 0}^1 |\phi(xy)\rangle \langle\phi(xy)| = \mathbb{I}_4$, if the set of unitary encoding chosen by Bob, $U^{xy}$, are mutually orthogonal\footnote{A set of unitary operators, $\{W_i\}_{i = 0}^{d^2 - 1}$, acting in a $d$-dimensional Hilbert space, $\mathcal{H}^d$, is called a complete set of orthogonal unitary operators, if it satisfies the orthogonality condition, given by
\(\frac{1}{d}\text{tr}(W_i W_j^\dagger) = \delta_{ij}\),
and the completeness relation
\(\frac{1}{d}\sum_i W_i \Xi W_i^\dagger = I_d\, \text{tr}\,\Xi\).
for some operator $\Xi \in \mathcal{H}^d$.} (for the proof see Appendix \ref{app:completeness-orthounitary}).
The Pauli matrices, along with the identity operator, can be one such example of mutually orthogonal unitary operators in the Hilbert space $\mathbb{C}^2$.  
Moreover, it is shown that for classical information transmission without security, with the help of a shared quantum state $\zeta_{AB}$ (the dense coding protocol), the maximal amount of information can only be sent, i.e., the dense coding capacity \cite{Bose2000, Ziman2003, DCBruss2006}, can be achieved when the encoding operators belong to a set of mutually orthogonal unitary operators and the probabilities are uniformly random~\cite{Hiroshima_2001}.


One can easily check that Eq.\eqref{eq:Bellwhat} is also a maximally entangled state, represented in a different basis choice in part of the sub-system $X$.  
The action of Eq. \eqref{eq:purification-general} can be considered as the state $\rho_{AX}$ concatenated with a maximally entangled state $|\phi^{+}\rangle_{X'B}$ 
followed by a generalized Bell measurement on the parties $XX'$, by using $|\phi(xy)\rangle$. Note that the initial state is product in the $X:X'$ bipartition, and the effect of different basis choice in subsystem $X$ is transferred as a unitary operation in part $B$, due to the effect of measurement, which is generally true in any teleportation or entanglement swapping operations.    
Each measurement outcome $\rho_{AB}^{xy}$ yields a particular encoded message $(x,y)$, which occurs with a uniformly random probability $\frac{1}{4}$. 

After the encoding process, Bob sends his qubit back to Alice through another quantum channel $\Lambda^b_{B \rightarrow A'}$, where the superscript `$b$' denotes the backward transmission. Alice's job is then to perform an optimal measurement $\mathcal{M}_{AA'} = \{M_{AA'}^{ij}\}$, over the two qubits in state $\rho_{AA'}^{xy}=\Lambda^b_{B \rightarrow A'}(\rho_{AB}^{xy})$, such that both the accessible information as well as the secret key rate are maximized. 

Thus, the probability with which Alice obtains the outcomes $(i,j)$, corresponding to Bob's encoding by $U^{xy}$, for some choice of $x,y$, is given by 
\begin{widetext}
\begin{eqnarray}
    p(ij|xy) &=& \text{tr} \left( M_{AA'}^{ij} \rho_{AA'}^{xy}(M_{AA'}^{ij})^\dagger\right) \\ &=& \text{tr} \left( M_{AA'}^{ij} \Lambda^b_{B \rightarrow A'}\left(\rho_{AB}^{xy}\right)(M_{AA'}^{ij})^\dagger\right) \\
    &=& \text{tr} \left( M_{AA'}^{ij} \Lambda^b_{B \rightarrow A'}\left(  4 \times {}_{XX'}\langle \phi(xy)|\left(\rho_{AX} \otimes |\phi^{+}\rangle\langle \phi^{+}|_{X'B}\right)|\phi(xy)\rangle_{XX'}  \right)(M_{AA'}^{ij})^\dagger\right). 
\end{eqnarray}
Considering $\Phi^{xy}_{XX'} = |\phi(xy)\rangle\langle \phi(xy)|_{XX'}$, and $p(ij;xy) = p(ij|xy) \times p(xy) = \frac 14 p(ij|xy)$, we get 
\begin{eqnarray}
    p(ij;xy) 
    =  \text{tr} \left( M_{AA'}^{ij}  \otimes  \Phi^{xy}_{XX'}\left(\rho_{AX} \otimes \Lambda^b_{B \rightarrow A'}\left(|\phi^{+}\rangle\langle \phi^{+}|_{X'B}\right) \right) (\Phi^{xy}_{XX'})^\dagger \otimes (M_{AA'}^{ij})^\dagger\right). 
\end{eqnarray}
From the above probability distribution, one can consider a post-measurement classical-classical ($cc$) state as $\sum_{ijxy}p(ij;xy) |ij\rangle\langle ij|_{AA'} \otimes |xy\rangle\langle xy|_{BB'}$, where we keep the subscript $BB'$, to ensure that it is Bob who keeps the classical register to store the bit values $(x,y)$, and Alice stores $(i,j)$.
\end{widetext}
One can easily check that in an ideal scenario, i.e., when both the $\Lambda^f_{A' \rightarrow B}$ and $\Lambda^b_{B \rightarrow A'} $ are noiseless, we should have a perfect correlation between Bob's encoding operation and Alice's measurement outcome, i.e., $p(ij|xy)= \delta_{ix}\delta_{jy}$. It automatically sets the choice of the optimal measurement in part of Alice as $M_{AA'}^{ij} = |\chi(ij)\rangle\langle \chi(ij)|_{AA'}$, (see Appendix \ref{app:proofdelta} for the proof), where the states $\{|\chi(ij)\rangle\}$ for SDC protocol are almost similar to  Eq. \eqref{eq:Bellwhat}, 
but with a small modification, 
\begin{equation}
     |\chi(ij)\rangle_{AA'} = \frac{1}{\sqrt{2}} \sum_{l,m=0}^{1} (U^{ij})_{ml}|l,m\rangle_{AA'} \label{eq:BellAlice}\,,
\end{equation}
where $\{U^{ij}\}$ are the same set of unitary operators used by Bob to encode his key bits, with $x$ replaced by $i$ and $y$ by $j$. For the measurement operators in the case of noise adaptive LM05 protocol, see Ref. \cite{Patra_PRA_2024_dimensional-advantage-SDC}, and Appendix \ref{app:LM05}.

It is important to note that a possible generalization of the set of mutually orthogonal unitary operators is the following:
$U^{xy}= W(\theta,\chi,\phi)\cdot \hat{\sigma}^{xy}$,  where $\hat{\sigma}^{xy}$ denotes the qubit identity and the Pauli matrices ($\hat{\sigma}^{00}=\mathbb{I},\, \hat{\sigma}^{01}=\hat{\sigma}^Z,\,\hat{\sigma}^{10}=\hat{\sigma}^X,\, \hat{\sigma}^{11}=-i \hat{\sigma}^Y$). Here, $W(\theta,\chi,\phi)$ denotes an arbitrary unitary operator in $SU(2)$ algebra, parametrized by 
\begin{equation}
     W(\theta,\chi,\phi) =
\begin{pmatrix}
\cos{(\frac \theta 2)}\, e^{\iota\chi} & \sin{(\frac \theta 2)}\, e^{\iota \phi} \\[0.5em]
-\sin{(\frac \theta 2)}\, e^{-\iota \phi}  &  \cos{(\frac \theta 2)}\, e^{-\iota \chi}
\end{pmatrix}\label{eq:W unitary},
\end{equation}
where $0 \leq \theta,\chi,\phi \leq \pi$. For this particular choice of the orthogonal unitary operators, the state $\rho_{AA'}^{xy}$ (see Eq. \eqref{eq:purification-general}, for $\rho_{AB}^{xy}$) can be written as 
\begin{widetext}
    \begin{eqnarray}
  \rho_{AA'}^{xy} &=& \Lambda^b_{B \rightarrow A'}\left(\rho_{AB}^{xy}\right) \nonumber \\
  &=& 4 \times {}_{XX'}\langle B(xy)|\left[\rho_{AX} \otimes \Lambda^b_{B \rightarrow A'}\left((\mathbb{I}_{X'}\otimes W_B) \,|\phi^{+}\rangle\langle \phi^{+}|_{X'B}\, (\mathbb{I}_{X'}\otimes W^{\dagger}_B)\right) \right] |B(xy)\rangle_{XX'}, \label{eq:orthogonal_purification}
\end{eqnarray}
where 
\begin{equation}
    |B(xy)\rangle = \frac{1}{\sqrt{2}} \sum_{l=0}^{1} e^{\iota\pi l y} |l,l\oplus x\rangle \label{eq Bell}\,
\end{equation}
are the four mutually orthogonal and maximally entangled Bell states corresponding to different values of $x,y \in \{0,1\}$.
The symbol ``$\oplus$'' denotes the addition modulo $2$.

The proof of Eq. \eqref{eq:orthogonal_purification} is given in Appendix \ref{app:proof_ortho_puri}. Let us now write $|\tilde{\phi}^{+}\rangle_{X'B} = (\mathbb{I}_{X'}\otimes W_B) \,|\phi^{+}\rangle_{X'B}$, and
$\tilde{\rho}_{X'A'} = \Lambda^b_{B \rightarrow A'}\left(|\tilde{\phi}^{+}\rangle\langle \tilde{\phi}^{+}|_{X'B}\right)$, hence the joint probability distribution can be written as
\begin{eqnarray}
    p(ij;xy) 
    =  \text{tr} \left( M_{AA'}^{ij}  \otimes  B^{xy}_{XX'}\left(\rho_{AX} \otimes \tilde{\rho}_{X'A'}\right) (B^{xy}_{XX'})^\dagger \otimes (M_{AA'}^{ij})^\dagger\right), \label{eq:prob_joint}
\end{eqnarray}
where $B^{xy}_{XX'} = |B(xy)\rangle\langle B(xy)|_{XX'}$.
\end{widetext}
To incorporate the effect of channel noise, we consider a generic $W(\theta,\chi,\phi)$, termed as the \emph{adaptive unitary}, and optimize the secret key rate in a one-way classical post-processing protocol \cite{Devetak_PRSA_2005_key-rate}, for different noise models, over the set of parameters $\theta,\chi,\text{and }\phi$.
Note that the state in Eq. \eqref{eq:Bellwhat} can now be represented  as 
\begin{equation}
    |\phi(xy)\rangle_{XX'} = \left(W_X (\theta,\chi,\phi) \otimes I_{X'}\right)|B(xy)\rangle_{XX'}, \label{eq:rotated-Bell}
\end{equation}
These orthogonal states, representing the measurement basis in $\mathbb{C}^2 \otimes \mathbb{C}^2$, can also be considered as the rotated Bell basis measurements.




\subsection{The modification of the secure key rate in NAQKD}\label{sec:generic-key-rate}
In this paper, we assume that Eve has the power to perform a collective attack, in which she can perform any operation allowed by the laws of quantum mechanics. In this scenario, Eve interacts with each shared state individually but stores her quantum systems (probes) in a quantum memory and postpones her measurement until Alice and Bob complete their classical post-processing. Eve can access all the information by performing a joint measurement on the $N$ copies of her stored quantum systems.

It is assumed that in the worst-case scenario, Eve holds the purifying system of the global state shared between the honest parties. The presence of an eavesdropping attack typically renders the initially shared entangled pure state between Alice and Bob mixed. This situation can also be interpreted as noise acting on the forward and backward quantum channels.

An equivalent description is that Eve prepares a global pure state and distributes the appropriate subsystems to Alice and Bob while retaining the purification. Let $|\psi\rangle_{AA'BB'E}$ denote the purification of the state $\rho_{AB} \otimes \tilde{\rho}_{A'B'}$\footnote{From this section onward, we will use the subscript $BB'$ to denote the subsystems of Bob. Note that the subscript $XX'$ is also with Bob.}, where $\rho_{AB}$ represents the state shared between Alice and Bob, and $\tilde{\rho}_{A'B'}$ corresponds to the auxiliary state required for the purification of Bob’s encoding operation. 

Now the classical–classical–quantum ($ccq$) state corresponding to a key-generation round, when both Alice and Bob perform their quantum operations, can be written as
\begin{equation}\label{eq:kappa-ccq}
    \begin{aligned}
        &\kappa_{AA'BB'E} =  \hat{\tilde{\mathcal{B}}}_{AA'} \otimes \hat{\mathcal{B}}_{BB'}(|\psi\rangle\langle \psi|_{AA'BB'E})\\
        &= \sum_{i,j,x,y} p(ij;xy)\, |ij\rangle\langle ij|_{AA'} \otimes |xy\rangle\langle xy|_{BB'} \otimes \rho_E^{ijxy} \,.
    \end{aligned}
\end{equation}

Here, $\hat{\tilde{\mathcal{B}}}_{AA'}(\cdot)$ (tilde) denotes the superoperator for Alice's measurement, given by 
\begin{eqnarray}
     \hat{\tilde{\mathcal{B}}}(\rho) &=& \sum_{i,j}|\chi(ij)\rangle \langle \chi(ij)| \rho |\chi(ij)\rangle \langle \chi(ij)| \\
     &\approx& \sum_{i,j} \langle \chi(ij)| \rho |\chi(ij)\rangle ~~|ij\rangle\langle ij| \label{eq:Bell-super}
\end{eqnarray}
for all $i,j \in \{0,1\}$ and $\ket{\chi(ij)}\approx \ket{ij}$\footnote{Here $M_{AA'}^{ij} = |\chi(ij)\rangle\langle \chi(ij)|_{AA'}$ takes this particular form for the SDC protocol, whereas for LM05, it is mentioned in Appendix \ref{app:LM05}.} being the rotated Bell measurement as in Eq.~\eqref{eq:BellAlice}. After the measurement, the quantum state collapses, and what remains is the information of the bit values $i$ and $j$. Thus, without loss of generality, Alice stores the information about the bit values $(i,j)$ in a classical register. For Bob's measurement, $\hat{\mathcal{B}}(\rho)$ can be similarly defined by the Bell states $\ket{B(xy)}$, with $\ket{B(xy)} \approx \ket{xy}$.

After completion of the quantum measurement by both Alice and Bob, the conditional quantum state in part of Eve reduces to $\rho_E^{ijxy}$, where $p(ij;xy)$ represents the joint probability distribution depending on the measurement statistics of Alice and Bob. Hence, Eve can extract information about the generated key by measuring her part of the shared state $\rho_E^{ijxy}$. The lower bound on the secure key rate, $r$, in case of the collective eavesdropping attack, when the honest parties perform a one-way classical post-processing protocol~\cite{Devetak_PRSA_2005_key-rate}, reads as
\begin{eqnarray}\label{eq:devetak-key}
    r &\geq& I(A:B)_{\kappa} - I(B:E)_{\kappa} \\
    &=& S(B|E)_{\kappa} - S(B|A)_{\kappa}\, ,\label{eq:devetak-key2}
\end{eqnarray}
where the mutual information\footnote{ $I(A:B)=S(\eta_A)+S(\eta_B)-S(\eta_{AB})$ with $S(\eta)=-\text{tr}(\eta \log_2 \eta)$ being the von Neumann entropy.} can be calculated with respect to the classical-classical-quantum ($ccq$) state given in Eq. \eqref{eq:kappa-ccq}. Remember that for a classical-classical ($cc$) state $\eta$, $I(A:B)_\eta = H(B)_{\eta}-H(B|A)_{\eta}$, 
with the von Neumann entropy $S(\eta_B)= - \text{tr}(\eta_B \log_2 \eta_B) = H(B)_{\eta}$ reduced to the Shannon entropy of the reduced density matrix $\eta_B = \text{tr}_A (\eta_{AB})$, and $S(B|A)_{\eta}= S(\eta_{AB}) - S(\eta_B) = H(AB)_{\eta}-H(A)_{\eta} = H(B|A)_{\eta}$, the  conditional Shannon entropy~\cite{Nielsen_2010_book}, for the $cc$ state. Eq. (\ref{eq:devetak-key}) computes the correlation between Bob and Alice versus Bob and Eve. If $I(B:E) > I(A:B)$, it implies that Eve has gained more information compared to Alice, and no final key can ever be distilled from the raw key, and hence the protocol needs to be aborted. 

Throughout the manuscript, we use state labels as the arguments of the entropy functions and denote the actual states by subscripts. To derive the secret key rate, let us consider the following two states:
\begin{align}
        &\xi_{AA'BB'E} = \hat{\tilde{\mathcal{G}}}_{AA'}\otimes \hat{\mathcal{B}}_{BB'} (|\psi\rangle\langle \psi|_{AA'BB'E}) \label{eq xi}, \\
        \text{and,}\quad &\tau_{AA'BB'E} = \hat{\tilde{\mathcal{G}}}_{AA'}\otimes \hat{\mathcal{G}}_{BB'}(|\psi\rangle\langle \psi|_{AA'BB'E}) \label{eq tau},
\end{align}
where $\hat{\mathcal{B}}$ is already given,
acting in the part of Bob,  whereas $\hat{\mathcal{G}}$ implies the test measurement performed by Bob to detect the presence of Eve, in the transmission channel, defined by 
\begin{eqnarray}
\hat{\mathcal{G}} (\rho) &=& \sum_{ij}|i_{\hat{n}},j_{\hat{m}} \rangle\,\langle i_{\hat{n}},j_{\hat{m}}|\,\rho\,|i_{\hat{n}},j_{\hat{m}} \rangle\,\langle i_{\hat{n}},j_{\hat{m}}| \label{eq:GBob} \\
&=& \sum_{ij} \langle i_{\hat{n}},j_{\hat{m}}|\,\rho\,|i_{\hat{n}},j_{\hat{m}} \rangle\, |ij\rangle \langle ij|
\end{eqnarray}
for all $i,j\in\{0,1\}$. The (product) bases $|i_{\hat{n}},j_{\hat{m}} \rangle$ are the eigenvectors of the rotated spin observable $\hat{n}\cdot\vec{\sigma} \otimes  \hat{m}\cdot\vec{\sigma}$. The superoperator $\hat{\tilde{\mathcal{G}}}_{AA'}(\cdot)$ (tilde) represents the noise-adaptive test measurement performed by Alice. Its action is the same as defined in \eqref{eq:GBob}, but the measurement basis is now changed to $| i_{\hat{n}},j_{\hat{m'}} \rangle \rightarrow (\mathbb{I}\otimes W(\theta,\chi,\phi))|i_{\hat{n}},j_{\hat{m'}} \rangle$\footnote{One can easily check that the eigenbasis of $\hat{m}\cdot\vec{\sigma}$, and $\hat{m'}\cdot\vec{\sigma}$ are unitarily connected. Suppose we choose $\hat{m} = \hat{z}$, then with the help of little algebra for any unitary $U = e^{i \theta \hat{n}.\vec{\sigma}}$, one can find  $U\hat{\sigma}_zU^\dagger = \hat{m'}.\vec{\sigma}$, where
\[
\hat{m'} =
\begin{pmatrix}
n_x n_z (1-\cos 2\theta) + n_y \sin 2\theta \\
n_y n_z (1-\cos 2\theta) - n_x \sin 2\theta \\
\cos 2\theta + n_z^2 (1-\cos 2\theta)
\end{pmatrix}.
\]
Here both $|\hat{n}| = |\hat{m'}| = 1$. For the purpose of maximal key rate, we choose the unitary operator to be $W(\theta,\chi,\phi)$, and optimize over the parameters $\theta,\chi,\phi$, for different noise models.}. 
 
 
For the two $ccq$ states $\kappa$ and $\xi$, one can readily check that $S(B|E)_{\kappa} = S(B|E)_{\xi}$, since the two states in Eqs.~\eqref{eq:kappa-ccq} and \eqref{eq xi} differ only in the measurements performed on Alice's subsystem, which does not affect the conditional von Neumann entropy between Bob and Eve.
 
In order to obtain the lower bound on Eve's knowledge about the secret key rate, we employ the entropic uncertainty relation formulated by \textit{Berta et al.}~\cite{Berta_NP_2010_uncertainty-principle-memory}. Suppose two different measurements are performed on the subsystem $A$ of a tripartite state $\rho_{ABE}$, using the POVM settings: $\Pi_X \equiv \{\Pi_X^i\}$ and $\Pi_Z \equiv \{\Pi_Z^j\}$, with classical outcomes denoted by $i$ and $j$, respectively. The corresponding entropic uncertainty relation states
\begin{equation}
    S(Z|B) + S(X|E) \geq \log_2 \frac{1}{\gamma},
\end{equation}
where $\gamma = \max_{i,j} || \sqrt{\Pi_X^i}\sqrt{\Pi_Z^j}||_\infty^{\,2}$, with $i$ and $j$ denoting the measurement outcomes of $\Pi_X$ and $\Pi_Z$, respectively, and $||.||_\infty$ stands for the infinity norm~\cite{Watrous_CUP_2018_book}. Applying this relation to the states in Eqs. (\ref{eq xi}) and (\ref{eq tau}), we obtain
\begin{equation}\label{eq ent unc B}
    S(B|E)_\xi + S(B|A)_\tau \geq \log_2 \frac{1}{\gamma}.
\end{equation}
Here $\gamma = \max_{(ij),(xy)} || \sqrt{\mathcal{G}(ij)}\sqrt{\mathcal{B}(xy)}||_\infty^{\,2}$. Using Eq. \eqref{eq:devetak-key2}, the lower bound on secure key rate becomes
\begin{eqnarray}\label{eq:working-key-rate}
        r &\geq&  S(B|E)_{\kappa}-S(B|A)_{\kappa} = S(B|E)_{\xi} - S(B|A)_{\kappa} \nonumber \\
        &\geq& \log_2 \frac{1}{\gamma} - S(B|A)_\tau - S(B|A)_{\kappa}.
\end{eqnarray}
If the measurement performed by Alice and Bob to obtain $\kappa_{AA'BB'E}$ and $\tau_{AA'BB'E}$ are fully correlated, which is the case when the shared state is maximally entangled, the last two von Neumann entropy terms in Eq. (\ref{eq:working-key-rate}) vanish and thus $r \geq \log_2 \frac{1}{\gamma}$. 

Each term in Eq.~\eqref{eq:working-key-rate} depends on the choice of the orthogonal unitary operators $U^{xy} = W(\theta,\chi,\phi)\,\hat{\sigma}^{xy}$, and hence on the parameters of the adaptive unitary $W(\theta,\chi,\phi)$. To achieve optimal performance for each noise model, the honest parties must choose those set of unitaries which maximize the key rate for that specific noise; hence the maximal noise-adaptive secret key rate should be: $\max_{\theta,\chi,\phi} \left( \log_2 \frac{1}{\gamma} - S(B|A)_\tau - S(B|A)_{\kappa}\right)$.
Moreover, if the noise in the transmission channel is very high, it is possible that the estimated key rate (given in Eq. \eqref{eq:working-key-rate}) by the honest parties computed by performing the key generation run and test run might turn out to be negative, in that case the honest parties abort the protocol, and then the key rate can be considered as $0$. Hence, the noise adaptive key rate can be defined as
\begin{eqnarray}
   \hspace{-0.5em}    \nonumber r_{adaptive} 
        &\geq& \max\bigg[0,\,\max_{\theta,\chi,\phi} \left( \log_2 \frac{1}{\gamma} - S(B|A)_\tau - S(B|A)_{\kappa}\right)\bigg],\\
        && \label{eq:working-key-rate-maximized}
\end{eqnarray}
where the optimization is performed over $\theta$, $\chi$, and $\phi$, involved in the unitary operator in Eq. \eqref{eq:W unitary}.

\section{Noise adaptive secure key rate under various uncorrelated noise models}\label{sec:results-noise-adaptive}

Let us calculate the $r_{adaptive}$ for various noise models, acting individually on the local subsystems possessed by Alice and Bob, by explicitly optimizing the parameters of the noise adaptive unitary operator $W(\theta,\chi,\phi)$. We now compare the noise adaptive key rate in Eq.~\eqref{eq:working-key-rate-maximized} for two fundamentally different two-way protocols, namely the superdense coding (SDC)~\cite{Bostrom_PRL_2002_ping-pong} and the LM05 protocols~\cite{Lucamracini_PRL_2005_LM05}, in the presence of various noise models within the noise-adaptive framework. In this setting, the forward and backward transmissions are modeled as identical but separate channels acting on the traveling qubit before and after the encoding process.
We quantify the advantage offered by the noise-adaptive strategy by comparing the resulting key rates with those of the corresponding standard (non-adaptive) protocols (for the detailed descriptions of the SDC and LM05 protocols and the respective key rates, see Appendices~\ref{app:SDC} and \ref{app:LM05}, respectively.).

In addition, we compare our results with a two-way variant of the BB84 protocol (see Appendix \ref{sec:secureBB84}). Although BB84 is inherently a prepare-and-measure protocol, we construct an effective two-way version by considering two consecutive BB84 rounds: in each cycle, one party prepares a quantum state, and the other performs a measurement, followed by a role reversal in the subsequent round. This pair of rounds is treated as a single effective two-way BB84 protocol for the purpose of comparison.

For independent noise, we identify regimes where no advantage is possible (e.g., depolarizing and bit flip channels), as well as scenarios where noise adaptation provides protocol-dependent improvements. These results signify the fact that, even in the presence of identical independent noise, an appropriate choice of state preparation and encoding can enhance the achievable key rate beyond conventional strategies.




\subsection{No-gain of NAQKD for paradigmatic uncorrelated noise models}{\label{A}}


\begin{figure*}[t]
    \includegraphics[width=0.8\linewidth]{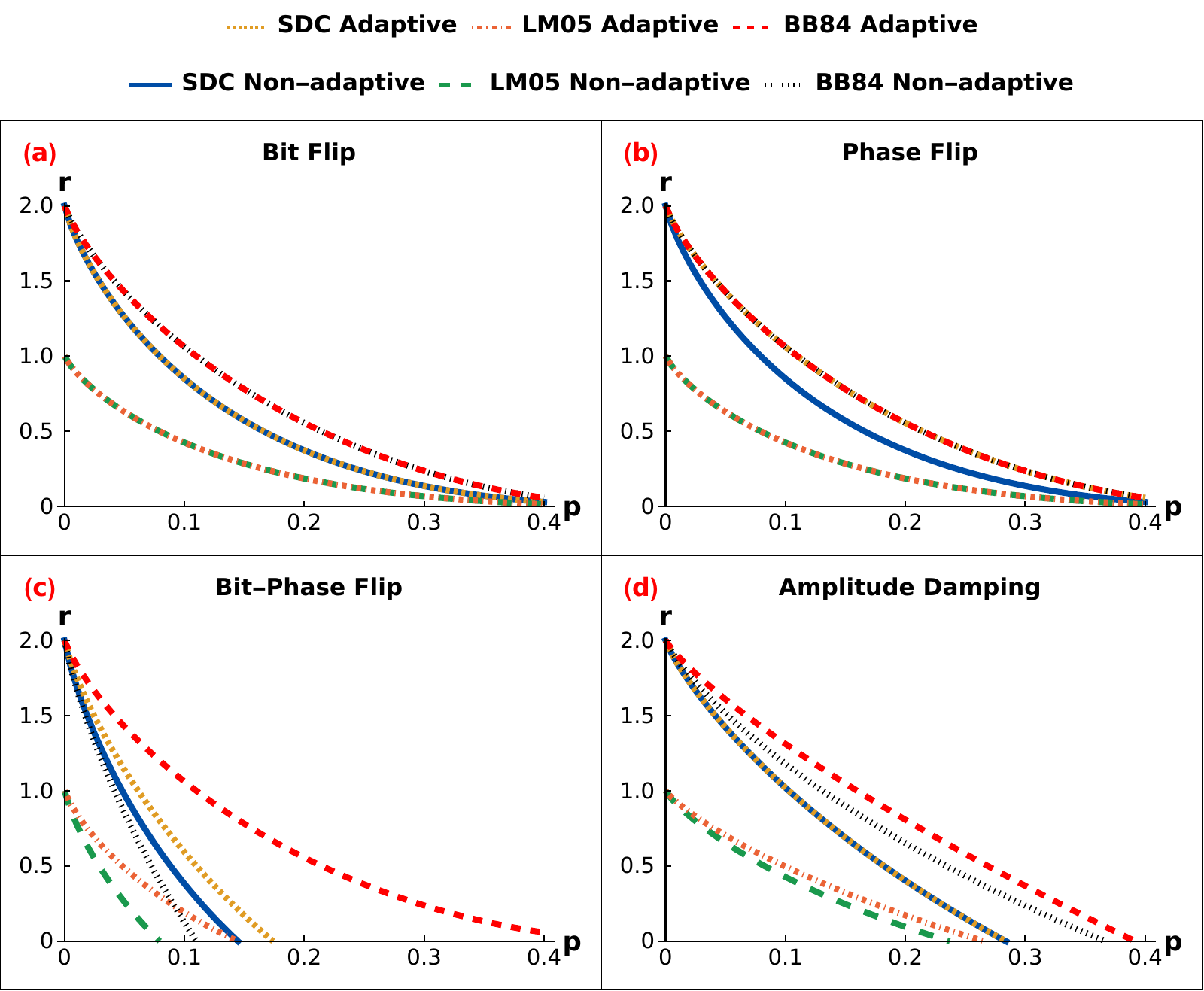}
    \caption{\justifying{(Color online) Key rate ($r$}) (ordinate) obtained with the adaptive and non-adaptive schemes against the noise strength ($p$) (abscissa) across four noise models (a) bit flip, (b) phase flip, (c) bit-phase flip, and, (d) amplitude damping channels, for all the three quantum key distribution protocols- secure dense coding (SDC), LM05 and two-way BB84. The independent noise acts with equal strength in both the forward and backward transmission channels. (a) For the bit flip channel, the optimal adaptive key rate and the non-adaptive key rate coincide for all three protocols over the whole range of $p$,  indicating no benefit from adaptive encoding (state preparation) and decoding (measurement) schemes. On the other hand, we observe the fact that for the phase flip channel ((b)), the adaptive key rate exceeds the non-adaptive key rate for the SDC protocol, although no advantage is observed for LM05 and two-way BB84 protocols. It is interesting to note that the adaptive key rate for the SDC protocol coincides with the key rate for the two-way BB84 protocol.  (c) depicts results for the bit-phase flip noise model, revealing improved key rates for all three QKD protocols, as well as enhancement of the critical noise threshold for secure information transmission, which is particularly significant for the 2-way BB84 protocol. (d) Similar findings are observed in the case of the local amplitude-damping channels for the LM05 and 2-way BB84 protocols, while the SDC protocol shows no improvement in this scenario. The noise adaptive key rate for the two-way BB84 protocol turns out to be the highest among all, for all four noise models. The horzontal axes are dimensionless while the vertical axis is in bits.}
    \label{fig:indi-all}
\end{figure*}

Let us first present the no-go theorems by proving that the adaptive scheme cannot provide any benefit over the conventional non-adaptive QKD schemes in the case of $SU(2)$ covariant and bit flip channels.

\textbf{Theorem 1: }\textit{For 
$SU(2)$ covariant channel, there is no improvement in secret key rate with the noise-adaptive protocol compared to the conventional scheme.} \\[3pt]
\textbf{Proof:} A quantum channel $\Lambda^c$ will be called $SU(2)$ covariant, if it satisfies:
\begin{equation}
    \Lambda^c(U \rho U^{\dagger}) = U \Lambda^c(\rho) U^{\dagger}
\end{equation}
where $U$ is an arbitrary unitary belonging to $SU(2)$.
Thus, the above relation also holds for the adaptive unitary $W(\theta,\chi,\phi)$. Therefore, for any $SU(2)$ covariant noisy channel $\Lambda^c$, we can write the encoding process by Bob, on $\rho_{AB}$ as
\begin{equation}
    \rho_{AB}^{xy} = U^{xy}_B\,\rho_{AB}\,U^{xy\,\dagger}_B = W_B\,\hat{\sigma}^{xy}_B\, \rho_{AB}\,\hat{\sigma}^{xy}_B\,W^{\dagger}_B
\end{equation}
since for the Pauli matrices $\hat{\sigma}^{xy\,\dagger}= \hat{\sigma}^{xy}$. The encoded state is then sent back to Alice through the channel $\Lambda^c_{B\rightarrow A'}$. Before being measured by Alice, the state can be expressed as
\begin{eqnarray}
    \rho_{AA'}^{xy}&=&\Lambda^c_{B\rightarrow A'}(\rho_{AB}^{xy}) \nonumber \\
    &=& \Lambda^c_{B\rightarrow A'} \left( W_B\,\hat{\sigma}^{xy}_B\, \rho_{AB}\,\hat{\sigma}^{xy}_B\,W^{\dagger}_B \right) \nonumber \\
    &=& W_{A'}\,\Lambda^c_{B\rightarrow A'} (\hat{\sigma}^{xy}_B\, \rho_{AB}\,\hat{\sigma}^{xy}_B)\,W^{\dagger}_{A'}
\end{eqnarray}
where the unitary $W_{B}$, previously acting on Bob's subsystem, now acts on the subsystem $A'$ of Alice, after passing through the covariant backward channel $\Lambda^b_{B \rightarrow A'} \equiv \Lambda^c_{B\rightarrow A'} $.
Now Alice measures the qubits $AA'$ in the rotated Bell basis defined in Eq. \eqref{eq:BellAlice}, which can also be expressed as $|\chi(ij)\rangle=(\mathbb{I}\otimes W(\theta,\chi,\phi))|B(ij)\rangle$, for $i,j \in \{0,1\}$, with $|B(ij)\rangle$ given in Eq. \eqref{eq Bell}. The conditional probability becomes
\begin{eqnarray}
    &&p(ij|xy) = {}_{AA'}\langle \chi(ij)|\rho_{AA'}^{xy}|\chi(ij)\rangle_{AA'} \nonumber \\
    &&\hspace{-2em} ={}_{AA'}\langle B(ij)|\,W_{A'}^\dagger\, W_{A'}\,\Lambda^c_{B\rightarrow A'} (\hat{\sigma}^{xy}_B\, \rho_{AB}\,\hat{\sigma}^{xy}_B)\,W^{\dagger}_{A'}\,W_{A'}|B(ij)\rangle_{AA'} \nonumber \\
    &=&{}_{AA'}\langle B(ij)|\,\Lambda^c_{B\rightarrow A'} (\hat{\sigma}^{xy}_B\, \rho_{AB}\,\hat{\sigma}^{xy}_B)\,|B(ij)\rangle_{AA'}\, , \label{eq:covarian-prob}
\end{eqnarray}
where we use $W^\dagger W =\mathbb{I}$. The last expression in Eq. \eqref{eq:covarian-prob} depicts that $p(ij|xy)$ is independent of $W(\theta, \chi, \phi)$. It means that the encoding operation with $U^{xy}$ followed by a measurement with $|\chi(ij)\rangle$ is basically the same as the Pauli encoding (with $\hat{\sigma}^{xy}$) followed by a standard Bell measurement, which is simply the non-adaptive SDC key generation protocol. Similarly, one can prove that the conditional probability distribution for the test run, in order to calculate the  $S(B|A)_{\tau}$ from $\tau_{AA'BB'E}$ (given in Eq. \eqref{eq tau}), is also independent of $W(\theta, \chi, \phi)$.
Thus, we conclude that for any $SU(2)$ covariant quantum channel, there is no improvement in the secret key rate from the noise-adaptive protocol, for any of the three key distribution protocols which we have considered.  \hfill $\blacksquare$


One such example of $SU(2)$ covariant channels is the depolarizing channel, characterized by an isotropic Pauli mapping, i.e., $\Lambda_{depol}(U \rho U^{\dagger}) = U \Lambda_{depol}(\rho) U^{\dagger}$. This channel uniformly shrinks the Bloch sphere towards the center, i.e., towards the maximally mixed state. Its action can be expressed as: $\rho \rightarrow \Lambda_{depol}(\rho) = p\frac{ \mathbb{I}}{2} + (1-p)\rho$, where $p$ denotes the noise strength. 




Let us now consider the case where both the forward and backward transmission channels are affected by bit flip noise. This is another noise model for which the adaptive protocol does not help. 

\textbf{Proposition 1:} \textit{No improvement in the secret key rate of NAQKD over the non-adaptive one is observed for the bit flip (BF) channel.}

The bit flip channel is characterized by the following Kraus operators
\begin{equation}
\label{eq 3}
    \Lambda^{BF}\; : \; \hat{K}_I = \sqrt{1-p}\,\mathbb{I}\,;\; \hat{K}_X = \sqrt{p}\,\hat{\sigma}^X
\end{equation}
We do not find any advantage in the secret key rate with the noise adaptive protocol for this channel, see Fig. \ref{fig:indi-all}(a), where we plot both the $r_{adaptive}$ and the conventional key rate $r$\footnote{Note that the conventional or the non-adaptive key rate can be obtained directly from Ineq. \eqref{eq:working-key-rate-maximized} by choosing $\theta = \chi = \phi =0$.} for all three quantum key distribution protocols, namely the SDC, LM05 and 2-way BB84 protocol. In all three cases, the adaptive key rates coincide with the conventional non-adaptive key rates. The key rates for SDC and BB84 start from $2$, whereas for LM05, they start from $1$. 
\begin{table*}[]
    \centering
    \begin{tabular}{|c|c|c|c|c|}
    \hline
    & \multicolumn{4}{c|}{ } \\[-0.5em]
    & \multicolumn{4}{c|}{ \bf Noise models} \\[0.5em]
    \hline
    \textbf{} 
 & \textbf{}
         & \multicolumn{2}{c|}{\cellcolor{red!30} \bf }
 & \textbf{} \\[-0.5em]
    \textbf{Protocol} 
 & \textbf{Phase flip}
         & \multicolumn{2}{c|}{\cellcolor{red!30} \bf Bit-phase flip}
 & \textbf{Amplitude damping} \\[0.5em]
 \hline
 \textbf{SDC} 
 & $\begin{matrix}
 \\[-0.5em]
  \left(\frac{\pi}{2},\chi,\chi\right), \\[0.5em]
  ~\frac{1}{\sqrt{2}}\begin{pmatrix}e^{i \chi}&e^{i \chi}\\-e^{-i \chi}&e^{-i \chi} 
  \end{pmatrix} ~
\vspace{0.5em}
\end{matrix}$
& $ \begin{matrix}
    \left(\pi,\chi,\frac{\pi}{4}\right), \\[0.5em]
    \frac{1}{\sqrt{2}}\begin{pmatrix}0 & 1+i\\-1+i & 0\end{pmatrix}
\end{matrix}$
 & $\begin{matrix}
     \left(\pi,\chi,\frac{3\pi}{4}\right), \\[0.5em]
     \frac{1}{\sqrt{2}}\begin{pmatrix}0 & -1+i\\1+i & 0\end{pmatrix}
 \end{matrix}$
  & $\mathbb{I}_2$\\
\cline{1-2} \cline{5-5}
\cellcolor{green!30}\textbf{LM05} 
& $\mathbb{I}_2$
& $\begin{matrix}
    \left(0,\frac{\pi}{4},\phi\right), \\[0.5em]
    \frac{1}{\sqrt{2}}\begin{pmatrix} 1+i & 0\\0 & 1-i\end{pmatrix} \vspace{0.5em}
\end{matrix}$
 & $\begin{matrix}
     \left(0,\frac{3\pi}{4},\phi\right), \\[0.5em]
     \frac{-1}{\sqrt{2}}\begin{pmatrix} 1-i & 0\\0 & 1+i\end{pmatrix} \vspace{0.5em}
 \end{matrix}$
& $\begin{matrix}
    \left(\frac{\pi}{2},\phi+\frac{\pi}{2},\phi\right), \\[0.5em] 
    \frac{1}{\sqrt{2}}\begin{pmatrix}
i e^{i\phi} & e^{i\phi}\\
-e^{-i\phi} & -i e^{-i\phi}
\end{pmatrix}
\end{matrix}$
 \\
\cline{1-1}
\cellcolor{blue!20}$~$ \textbf{2-way BB84} $~$
& 
& $\begin{matrix}
    \left(\theta,\frac{\pi}{4},\frac{3\pi}{4}
\right), \\[0.5em] 
~\begin{pmatrix}\frac{1+i}{\sqrt{2}}\,\cos{\frac{\theta}{2}} & \frac{-1+i}{\sqrt{2}}\,\sin{\frac{\theta}{2}} \\\frac{1+i}{\sqrt{2}}\,\sin{\frac{\theta}{2}} & \frac{1-i}{\sqrt{2}}\,\cos{\frac{\theta}{2}}\end{pmatrix} ~
\end{matrix}$
&  $\begin{matrix}
    \left(\theta,\frac{3\pi}{4},\frac{\pi}{4}\right), \\[0.5em]
~ \begin{pmatrix}\frac{-1+i}{\sqrt{2}}\,\cos{\frac{\theta}{2}} &\frac{1+i}{\sqrt{2}}\,\sin{\frac{\theta}{2}}\\\frac{-1+i}{\sqrt{2}}\,\sin{\frac{\theta}{2}} &-\frac{1+i}{\sqrt{2}} \,\cos{\frac{\theta}{2}}\end{pmatrix} ~
\end{matrix}$
& $\begin{matrix}
    \left(\frac{\pi}{2},\chi,\chi+\frac{\pi}{2}\right), \\[0.5em] 
    \frac{1}{\sqrt{2}}\begin{pmatrix}
e^{i\chi} & i e^{i\chi}\\
ie^{-i\chi} & e^{-i\chi}
\end{pmatrix} 
\end{matrix}$
\\
\hline
    \end{tabular}
    \caption{\justifying{
 This table presents the optimal parameters ($\theta$,$\chi$,$\phi$) and the corresponding adaptive unitaries $W$ for various protocols under phase flip, bit-phase flip, and amplitude-damping noise models. In case of phase flip noise, we find advantage only for the SDC protocol, with optimal unitary $\hat{\sigma}^Z\mathbf{H}$ $(\chi=\phi=0)$. On the other hand, under bit-phase flip noise action, we observe an improvement in secret key rate for all three protocols, with the optimal unitary operators being identical across each protocol. In total, six types of optimal unitaries achieve the same maximized key rate in the adaptive scenario. For the amplitude-damping channel, the SDC protocol does not provide any advantage; however, we do see an enhancement in the secret key rate for both the LM05 and 2-way BB84 protocols. The optimal unitaries are the same for both of these protocols and are listed above.}}
 \label{table:unitaries}
\end{table*}

\subsection{Enhancement of secret key rate with adaptive scheme}\label{sec: independent improvement}
Let us discuss here two noise models, namely the phase flip and the bit-phase flip noise, which again act independently on the forward and backward transmission channels. These two noise models are -- distinct, since one of these shows ubiquitous advantage for the noise-adaptive scheme, while the other one shows advantage only for a single QKD protocol.

\textbf{Proposition 2:} \textit{In the case of phase flip (PF) channel, the noise-adaptive protocol enhances the secret key rate for the SDC protocol, but no benefit can be observed for the LM05 and two-way BB84 protocol compared to non-adaptive ones.}\\[3pt]
The Kraus operator decomposition for the qubit phase flip channel reads
\begin{equation}
\label{eq dephase}
    \Lambda^{PF}\; : \; \hat{K}_I = \sqrt{1-p}\,\mathbb{I}\,;\; \hat{K}_Z = \sqrt{p}\,\hat{\sigma}^Z .
\end{equation}
The noise-adaptive protocol is found to increase the secret key rate over the whole range of the noise parameter $p$, for the SDC protocol. However, the criticality of $p$ ($p_{cr}$, the value of $p$ after which the key rate vanishes, implying no secure transmission of information) remains the same as that of the standard (non-adaptive) SDC protocol. The result can be seen from Fig. \ref{fig:indi-all}(b).
In this case, the optimal adaptive unitary, which makes the secret key rate maximum, is obtained with the combinations: $\theta\,=\pi/2\,\text{ and } \chi = \phi$. If we choose $\chi = \phi = 0$, it reduces to $\hat{\sigma}^Z.\mathbf{H}$ (where $\mathbf{H}$ represents the Hadamard unitary), which belongs to the Clifford group~\cite{clifford_PhysRevA.57.127}. Hence, for the SDC protocol under independent phase flip noise action, the optimal adaptive unitaries are
\begin{equation}
     W_{\text{optimal}}^{PF} = \frac{1}{\sqrt{2}}
\begin{bmatrix}
e^{i \chi} & e^{i \chi} \\
-e^{-i \chi} & e^{-i \chi}
\end{bmatrix} .
\end{equation}
For example, when $\theta=\frac{\pi}{2} \text{ and } \chi=\phi=0$,
\begin{equation}
    W_{\text{optimal}}^{PF} = \hat{\sigma}^Z.\mathbf{H} = \frac{1}{\sqrt{2}}
\begin{bmatrix}
1 & 1 \\
-1 & 1
\end{bmatrix}.
\end{equation}
It is interesting to observe that although there is no advantage for the LM05 as well as the 2-way BB84 protocol, the 2-way BB84 protocol provides the highest key rate over the entire range of $p \in [0,0.5]$ compared to the other two protocols. Additionally, the increased key rate from the adaptive SDC protocol coincides with the key rate of the BB84 protocol.\\[2pt]

\textbf{Proposition 3:} \textit{In the case of bit-phase flip (BPF) channel, the noise-adaptive protocol enhances the secret key rate for all three adaptive quantum key distribution protocols in comparison with the conventional ones.}\\[3pt]
The qubit bit-phase flip noise model is characterized by the Kraus operators
\begin{equation}
\label{eq bitphase}
    \Lambda^{BPF}\; : \; \hat{K}_I = \sqrt{1-p}\,\mathbb{I}\,;\; \hat{K}_Y = \sqrt{p}\,\hat{\sigma}^Y.
\end{equation}
In the case of the bit-phase flip channel, the noise-adaptive framework not only increases the secret key rate over the whole range of noise parameter $p$, but also increases its critical value, for all three noise-adaptive QKD protocols (see Fig. \ref{fig:indi-all}(c)). It clearly establishes the importance of the adaptive protocol to achieve security.
It is interesting to note that, contrary to the bit flip and phase flip noise, the conventional key rate for the 2-way BB84 protocol is lower than the conventional non-adaptive key rate for the SDC protocol. But in the case of the noise adaptive protocol with key rate $r_{adaptive}$, we observe that it increases significantly as compared to the case of the adaptive SDC protocol, such that the optimal adaptive key rate as well as the critical noise parameter become much higher for the 2-way BB84 protocol. 

Here, we obtain several optimal adaptive unitaries that are equally useful in providing a higher key rate compared to the non-adaptive ones, thereby indicating its nonuniqueness. For example, the angle combinations are: $(\theta,\chi,\phi) = (0,\,\frac{\pi}{4} \text{ or }\frac{3\pi}{4},\,\phi)$, $(\pi,\, \chi,\, \frac{\pi}{4}\text{ or }\frac{3\pi}{4})$, $(\theta, \frac{\pi}{4}, \frac{3\pi}{4})$ or $(\theta,\frac{3\pi}{4},\frac{\pi}{4})$. Notice that all combinations, except for the last two, again belong to the Clifford group~\cite{clifford_PhysRevA.57.127}. 
The list of all possible optimal unitary operators associated with the different optimal $(\theta,\chi,\phi)$, that give rise to the same maximized $r_{adaptive}$, is presented in Table \ref{table:unitaries}.

Beyond Pauli noise, if one considers the action of uncorrelated non-Pauli noise, the benefit of the adaptive scheme over various two-way QKD protocols is not universal, as already seen in the case of PF and BPF noise.\\

\textbf{Proposition 4:} \textit{The noise adaptive protocol enhances the key rate for LM05 and 2-way BB84 protocol, while no enhancement can be found for SDC protocol, when both the forward and backward transmission channels are affected by independent amplitude damping (AD) noise.}\\[3pt]
The Kraus operator decomposition of the amplitude damping channel, represented in the computational basis, can be written as
\begin{equation}
\label{eq:ampdamp}
    \Lambda^{AD} \; : \; \hat{K}_1 = |0\rangle\langle0| + \sqrt{1-p}\,|1\rangle\langle1| \,;\, \hat{K}_2 = \sqrt{p}\,|0\rangle\langle1|\,.
\end{equation}

\begin{figure*}[t]\label{fig:general pauli sdc}
\centering 
\begin{subfigure}{0.32\textwidth} 
\includegraphics[width=\linewidth]{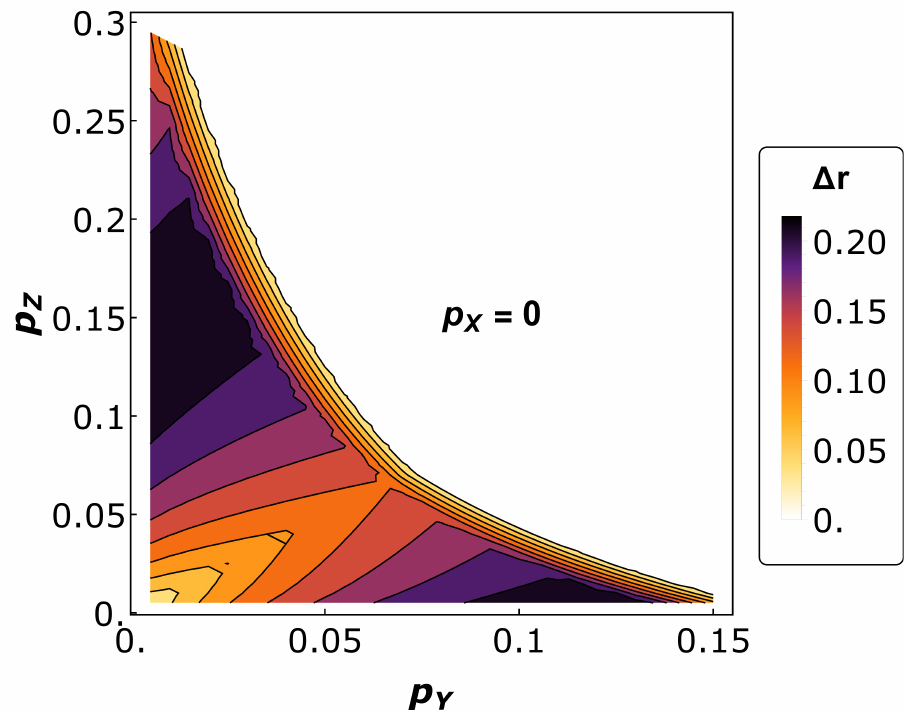}
\caption{}
\end{subfigure}
\hspace{5pt}
\begin{subfigure}{0.32\textwidth} 
\includegraphics[width=\linewidth]{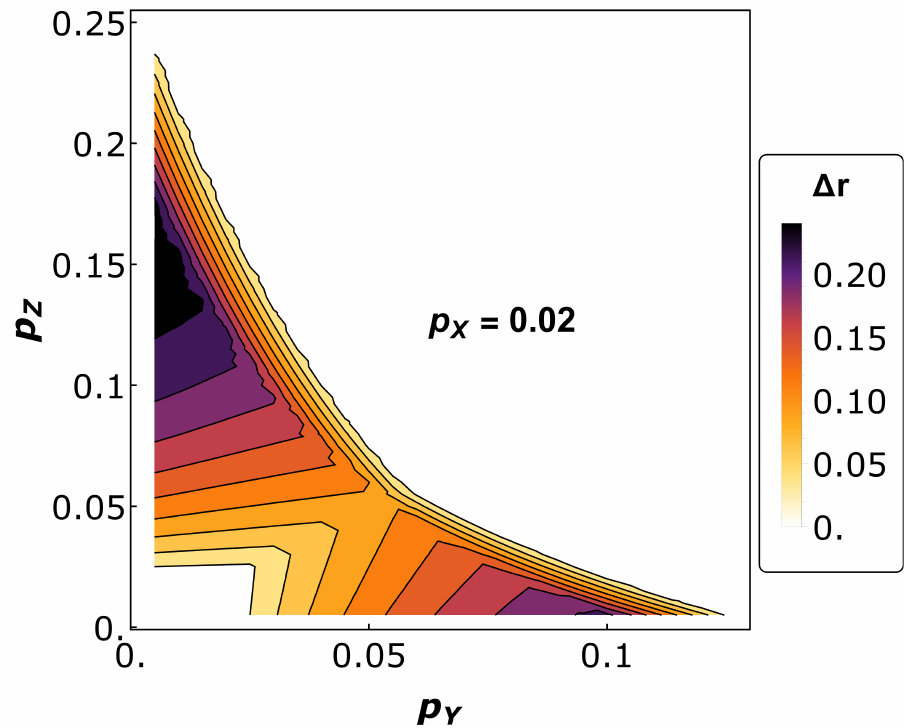}
\caption{}
\end{subfigure} 
\hspace{5pt}
\begin{subfigure}{0.32\textwidth} 
\includegraphics[width=\linewidth]{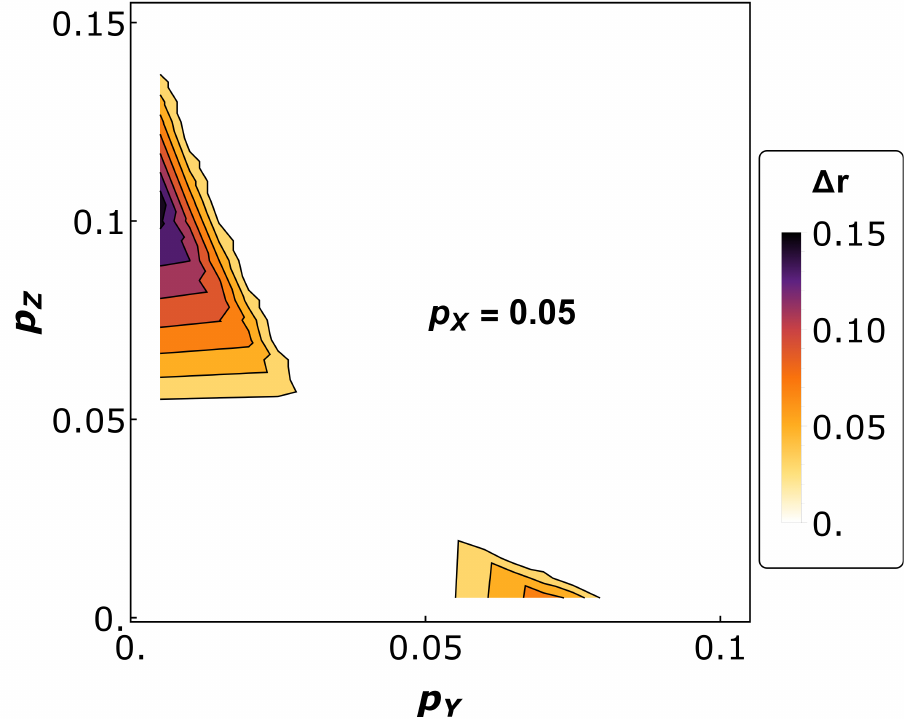}
\caption{}
\end{subfigure}
\caption{\justifying{(Color online)}{Map plot of difference in the key rate between the noise-adaptive and the conventional protocol ($\Delta r$ in Eq. (\ref{eq:differenceinkeyrates}))  for the SDC protocol under a general Pauli noise channel as a function of noise parameters $p_Y$ (abscissa) and $p_Z$ (ordinate), keeping fixed values of $p_X$. Here, the parameters $p_X$, $p_Y$, and $p_Z$ respectively represent the contributions from bit, bit-phase, and phase flip noises. (a) $p_X=0$,  (b) $p_X=0.02$, and (c) $p_X=0.05$. Note that with increasing contributions from bit flip noise elements, i.e.e, \(p_X\), we obtain less advantage from the adaptive protocol in the case of a general Pauli channel.}}
\label{fig:general pauli} 
\end{figure*}

\noindent In this case, the adaptive key rate $r_{adaptive}$, as well as the critical noise strength, for both the LM05 and the 2-way BB84 protocols, are higher than the conventional schemes. However, the adaptive SDC protocol does not provide any gain in key rate (see Fig.\ref{fig:indi-all} (d)). 

The optimal adaptive unitaries are the same for both the LM05 and the 2-way BB84 protocols. Two different sets of optimal unitaries are obtained, which work equally well to improve the secret key rate under the noise adaptive approach. The angle combinations and the associated unitary operators are as follows:\\[1pt]
\begin{equation}
    \left(\theta=\frac{\pi}{2}, \phi=\chi+\frac{\pi}{2} \right):\; W_{\text{opt\,1}}^{ad}=\frac{1}{\sqrt{2}}
\begin{bmatrix}
e^{i \chi} & ie^{i\chi} \\
ie^{-i \chi} & e^{-i \chi}
\end{bmatrix},
\end{equation}
\begin{equation}
    \hspace{-1em}\text{and, } \left(\theta=\frac{\pi}{2}, \chi=\phi+\frac{\pi}{2} \right) \;W_{\text{opt\,2}}^{ad}=\frac{1}{\sqrt{2}}
\begin{bmatrix}
ie^{i \phi} & e^{i\phi} \\
-e^{-i \phi} & -ie^{-i \phi}
\end{bmatrix}.
\end{equation}

\noindent \textbf{Remarks:} In all the noise models, the adaptive two-way BB84 protocol always provides the highest key rate among all the noise-adaptive and non-adaptive two-way QKD protocols.

\subsubsection{General Pauli channel}

Having established positive results for specific Pauli noise models across all three protocols, we proceed to analyze the most general case, where all three Pauli operators, and the qubit identity, act upon the input state with random probabilities $p_i \in [0,1]$~\cite{general_pauli_PhysRevA.105.032435}. The corresponding probabilities are constrained by the complete positive and trace-preserving (CPTP) condition~\cite{z_shadman_PhysRevA.84.042309}. This type of channel shows a mixed effect of bit flip, phase flip, and bit-phase flip operations, depending upon the probabilities of their occurrence. The channel action on any arbitrary state $\rho$ can be expressed in the operator-sum representation as
\begin{equation}
    \rho \rightarrow \Lambda^{Pauli}(\rho) = p_0 \rho +  \sum_{i=X,Y,Z} p_i\,\hat{\sigma}^i\rho\,\hat{\sigma}^i,
\end{equation}
with $\sum_{i=X,Y,Z} p_i = 1 - p_0$.

In this case, when a specific Pauli noise, such as phase flip or bit-phase flip, acts with a significantly higher probability compared to the others, the channel behaves accordingly, reproducing similar results governed by the same optimal adaptive unitaries. But when the contributions from different noise elements are nearly equal, the adaptive protocol yields only marginal improvements in the key rate with different types of optimal unitaries. We observe, however, that most of the optimal unitaries follow from the Clifford group. As discussed earlier, we do not get any advantage from the adaptive protocol for a pure bit flip noise channel, in any of the protocols. However, when combined with the effects of phase flip and bit-phase flip noise, it yields a higher key rate than the non-adaptive protocol.
To illustrate the advantage of the adaptive protocol, we introduce the quantity
\begin{equation}
    \Delta r = r_{adaptive} - r_{conventional},
    \label{eq:differenceinkeyrates}
\end{equation}
where $r_{adaptive}$
and $r_{conventional}$ denote the secret key rates obtained through the optimal adaptive and the conventional schemes, respectively, and we study its behavior as a function of noise parameters $p_Y$ and $p_Z$ for various fixed values of $p_X$, for the SDC protocol (see Fig. \ref{fig:general pauli}). Note that $p_Y$ and $p_Z$ represent the corresponding weightage of bit-phase flip and phase flip noises, whereas $p_X$ represents the contribution of bit flip noise in the system.

In all the cases, we can observe non-vanishing values of $\Delta r$, which indicate the enhancement in secret key rate by the noise-adaptive protocol. We note that the gain in key rate is marginal when the contributions of $p_Y$ and $p_Z$ in the system are very small. These can be seen by the increasing areas of white regions in Figs.~\ref{fig:general pauli}(b) and \ref{fig:general pauli}(c), near the origin. This is due to the fact that when the values of $p_Y$ and $p_Z$ lie within the range, say $[0,0.025)$, for Fig.~\ref{fig:general pauli}(b), and $[0,0.05)$, for Fig.~\ref{fig:general pauli}(c), the difference in key rate, $\Delta r \approx 0$, as in this region $r_{adaptive} \approx r_{conventional}$. However, in Fig.~\ref{fig:general pauli}(a), where $p_X = 0$, except for the point $p_Y = p_Z = 0$, an advantage in the key rate emerges for very small values of $p_Y$ and $p_Z$. Similar observations can be seen for high values of $p_Y$ and $p_Z$.

In most of the cases, we obtain that the optimal unitary turns out to be $(\frac{\pi}{2}, \frac{\pi}{4}, \frac{\pi}{4})$ or $(\frac{\pi}{2}, \frac{3\pi}{4}, \frac{3\pi}{4})$ or $(\frac{\pi}{2}, \frac{\pi}{2}, \frac{\pi}{2})$, which basically follow from the set $(\frac{\pi}{2},\chi,\chi)$ listed in Table~\ref{table:unitaries}, for phase flip noise in the SDC case. In the rest of the cases, the optimal unitary is one of the six types of unitaries (e.g., $(0,\frac{\pi}{4},\phi)$ or $(\pi,\chi,\frac{3\pi}{4})$), listed for the bit-phase flip noise case in Table~\ref{table:unitaries}. 

\section{Impact of adaptive protocol in correlated noise}
\label{sec:correlated-channel}

This situation involves the same noise action occurring both before and after the encoding process on the traveling qubit. This is characterized by the existence of correlations between successive applications of the channel~\cite{Macchiavello_PhysRevA.65.050301, Dynamical_memory_Macchiavello}. 
We first prove again a no-gain scenario when fully correlated noise influences the forward and backward transmission channels, and then we present the beneficial situation of the NAQKD scheme. 

In the case of two consecutive uses of a channel (for forth-and-back travel of the qubit in a two-way protocol, constituting any particular run), the combined channel action is given by
\begin{equation}
    \rho \rightarrow \Lambda(\rho) = \sum_{i, j} p_{ij} (\hat{K_i} \otimes \hat{K_j}) \,\rho\, (\hat{K_i}^\dagger \otimes \hat{K_j}^\dagger),
\end{equation}
where $p_{ij}$ is the joint probability distribution corresponding to the consecutive actions of Kraus operators $\hat{K_i}$, characterizing the noise model. In the general case of partially correlated channels, $p_{ij}$ takes the form,
\begin{equation}
    p_{ij}=(1-\mu)p_i p_j + \mu p_i \delta_{ij}\, ,
\end{equation}
where $p_i$ is the probability corresponding to the noise elements $\hat{K_i}$. The degree of classical correlations is characterized by $\mu\in [0,1]$, which, with some probability, forces the same Kraus operator to be applied in the consecutive use of the transmission channel. For $\mu=0$, we arrive at the uncorrelated or independent noise action scenario (Sec. \ref{sec:results-noise-adaptive}), where the joint probability gets factorized, i.e., $p_{ij} = p_i\, p_j$. The channel actions become fully correlated for $\mu = 1$, leading to the joint probability $p_{ij} = p_i \delta_{ij}$, which ensures that the same Kraus operator is applied with certainty in both uses of the channel (Sec. \ref{fully correlated}) while for $0<\mu<1$ (partially correlated scenario), the results are presented in Sec. \ref{partially correlated}. \\


\subsection{No-gain  for fully correlated Pauli noise channels}{\label{fully correlated}}
\textbf{Proposition 5:} \textit{Under fully correlated Pauli noise acting on the forward and backward transmission channels, the secret key rates of the noise-adaptive and non-adaptive  QKD protocols become identical, thereby establishing the absence of any advantage for the NAQKD scheme.}

\textbf{Proof: }In the forward channel, any Pauli noise $\sigma^i$ acts with some probability $p_i$, then it is guaranteed that in the backward channel, the same noise element $\sigma^i$ acts with unit probability. Hence, the resultant state prior to the measurement by Alice is given by
\begin{equation}{\label{corellated 1}}
    \rho_{AA'}^{xy} = \sum_i p_i\, \hat{\sigma}^i_{A'} \left( W_{A'}\,\hat{\sigma}^{xy}_{A'} \left(\hat{\sigma}^i_{A'}\, \rho_{AA'}\, \hat{\sigma}^i_{A'} \right)\hat{\sigma}^{xy}_{A'}\, W_{A'}^{\dagger}     \right) \hat{\sigma}^i_{A'}
\end{equation} \\
where, for simplicity, we ignore the suffix in the transmission channels $A'\rightarrow B$ and $B \rightarrow A'$, and the adaptive encoding operators are taken as $U^{xy}=W\,\sigma^{xy}$. 
In case the adaptive unitary $W$ is considered to be $\mathbb{I}$ (which is the case for the non-adaptive scenario), one can easily find that a perfect correlation between Bob's encoding bits $(x,y)$ and Alice's decoding bits $(i,j)$ can be obtained. This is because of the fact that for Pauli operators
\begin{equation}
    \hat{\sigma}^i\, \hat{\sigma}^k\, \hat{\sigma}^i =
\begin{cases}
\hspace{0.8em}\hat{\sigma}^i & \text{if } i = k \\
-\hat{\sigma}^k & \text{if } i \ne k\;.
\end{cases}
\end{equation}
Hence, we have 
\begin{equation}{\label{corellated 2}}
    \rho_{AA'}^{xy} = \sum_i p_i\,\hat{\sigma}^{xy}_{A'}\, \rho_{AA'}\, \hat{\sigma}^{xy}_{A'} = |B(xy)\rangle \langle B(xy)|_{AA'}
\end{equation}
Now, at the decoding step, Alice uses the Bell measurement $\hat{\mathcal{B}}^{\,ij}_{AA'}$, for optimal information gain, which results in any of the Bell states (denoted by $(ij)$), with unit probability, depending upon the bit values $(xy)$. As a result of this perfect correlation established in the key generation run, the secure key rate becomes much higher compared to the independent noise case.\\[2pt]
We observe numerically that, with the noise-adaptive protocol, the key rate for fully correlated noise does not improve anymore; it stays the same as the non-adaptive rate, returning the optimal adaptive unitary to be $W_{opt}=\mathbb{I}$. \hfill $\blacksquare$

\begin{figure*}[t]
    \centering
    \includegraphics[width=\textwidth]{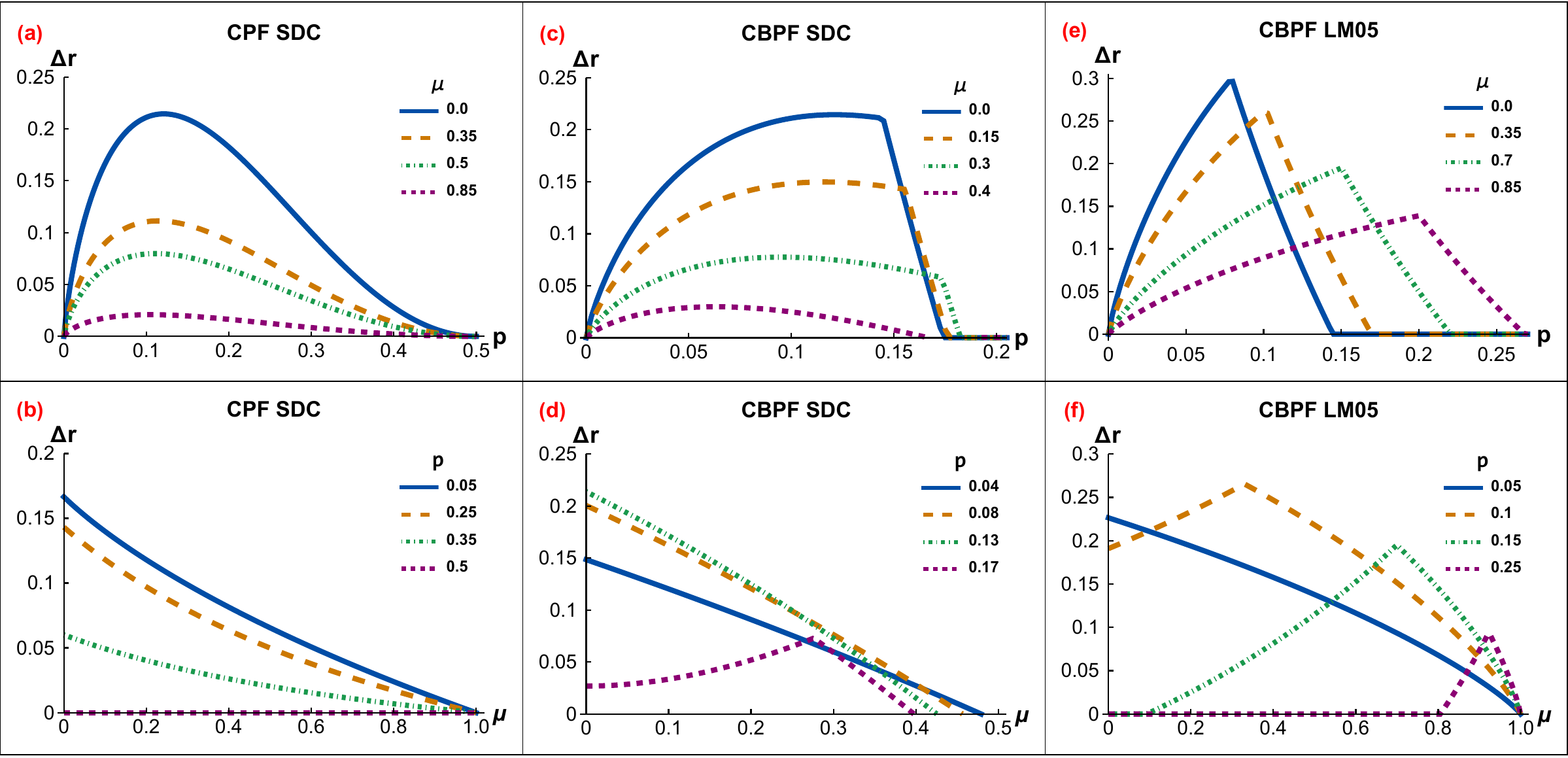}
    \caption{\justifying{(Color online) Difference in key rate of adaptive and non-adaptive protocols, $\Delta r$ (ordinate), against the noise strength $p$ and the degree of correlations $\mu$ (abscissa). The upper panel shows $\Delta r$ versus $p$ for fixed values of $\mu$, while the lower panel shows $\Delta r$ versus $\mu$ for fixed values of $p$. Figures $(a)$ and $(b)$ correspond to the SDC protocol under partially correlated phase flip noise, figures $(c)$ and $(d)$ to the SDC protocol under partially correlated bit-phase flip noise, and figures $(e)$ and $(f)$ to the LM05 protocol under partially correlated bit-phase flip noise. 
    }}
    \label{fig:partial-correlated}
\end{figure*}

\subsection{Beneficial role of adaptive QKD in partially correlated noise}{\label{partially correlated}}

\textbf{Proposition 6:} \textit{The adaptive protocol provides an advantage for the SDC protocol under partially correlated bit-phase flip and phase flip noise models, while for the LM05 protocol, such an advantage arises only in the former noise model.}

For partially correlated bit-phase flip noise (CBPF), we obtain an improvement in the secret key rate through the adaptive scheme for both the SDC and the LM05 protocols over the range of $p$ and $\mu$, while such an advantage can also be observed for partially correlated phase flip noise (CPF), only for the SDC protocol. Let us first elaborate on some of the observations for the SDC protocol:

$(1)$ \textit{$p$-dependence: }For the CBPF case, the key rate decreases with the increase in the noise strength $p$ and vanishes after a certain threshold ($p_{cr}$), for all values of $\mu$. By using the adaptive protocol, one can see a clear advantage in the secret key rate, as well as an increase in $p_{cr}$ with the adaptive scheme (see Fig. \ref{fig:partial-correlated}(c)). On the other hand, in the case of CPF, although there is no critical value of $p$ for all $\mu\geq 0$ (i.e., secure transmission over all noise strength and all degrees of correlation), $\Delta r>0$  however, in the entire region of $p$ and \(\mu\), thereby confirming the importance of the adaptive method over the non-adaptive ones (see Fig. \ref{fig:partial-correlated}(a) and (b)). 

$(2)$ \textit{$\mu$-dependence: }When the CBPF is present in the channels, the difference between $r_{adaptive}$ and $r_{conventional}$ decreases with increasing $\mu$, although in the critical $p$ region (e.g. $0.15 <p<0.2$), we observe a discrepancy. This is because for high $p$ (i.e., $p\approx p_{cr}$) and low $\mu$, $r_{conventional}$ can vanish, while $r_{adaptive}>0$. For instance, when $\mu < 0.275$,  $r_{conventional} =0$, while $r_{adaptive}>0$, for $p=0.17$, after which both the key rates become positive, resulting in kink in the curve (see Fig. \ref{fig:partial-correlated}(d)). 

\begin{figure*}[t]
    \includegraphics[width=0.8\linewidth]{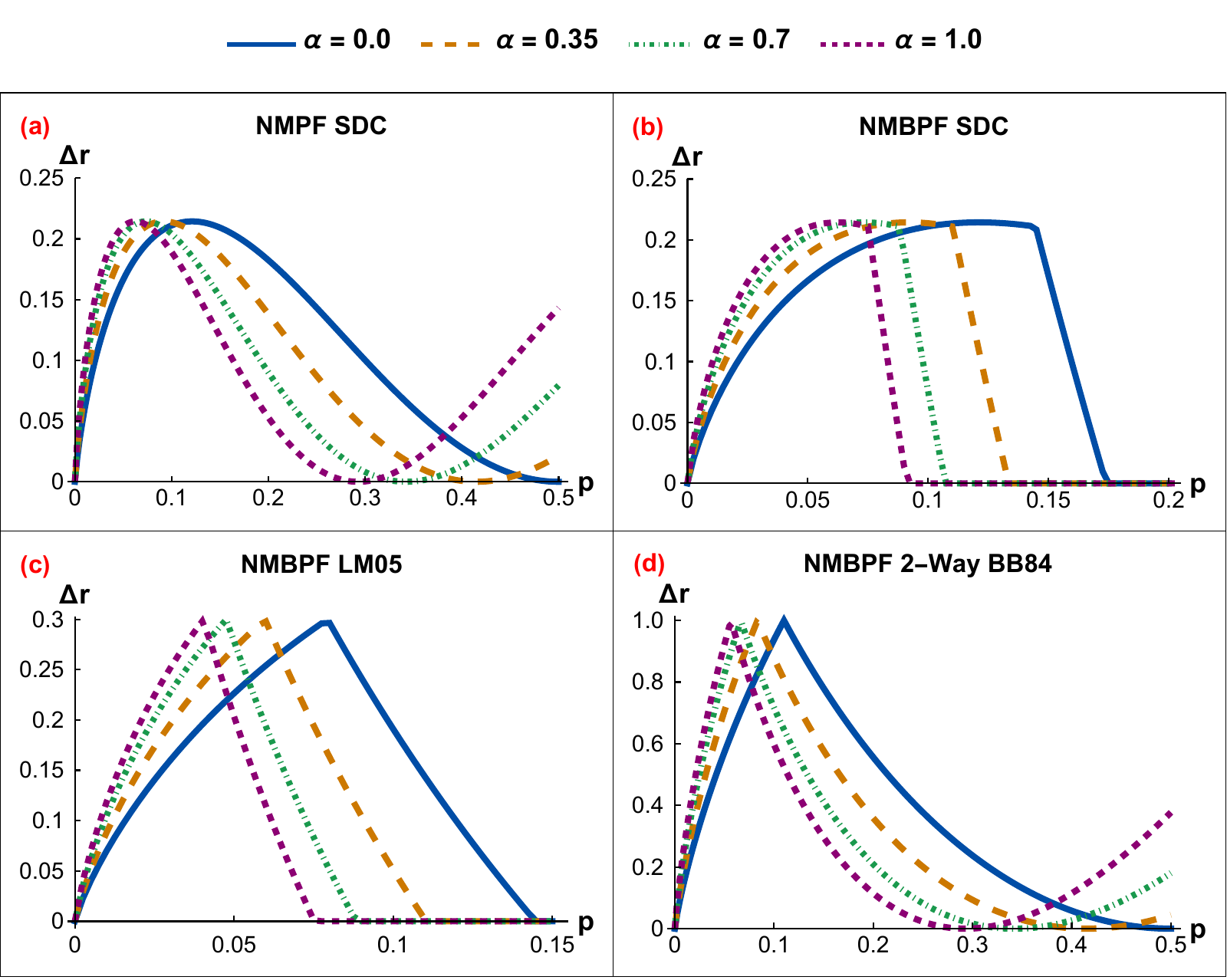}
    \caption{\justifying{(Color online) 
    $\Delta r$ (ordinate) with respect to $p$ (abscissa) for fixed values of the non-Markovianity parameter $\alpha$. (a) and (b)  the SDC protocol under NM phase flip noise (a), and NM bit-phase flip noise (b). Under NM bit-phase flip noise, in (c), \(\delta r\) is obtained for the LM05 protocol, while in (d), \(\delta r\) is computed for the two-way BB84 protocol. In each of the plots, the (blue) solid lines correspond to $\alpha=0$,  long-dashed, dot-dashed, and the short-dashed lines represent the increase of Non Markovianity from $\alpha=0.35$, $\alpha=0.7$, to $\alpha=1$.}}
    \label{fig:nonmark-all}
\end{figure*}

For the CPF case, $\Delta r$ decreases monotonically with increasing $\mu$, for all values of $p$, although over the entire range of $\mu$, $\Delta r>0$ (see Fig. \ref{fig:partial-correlated}(b)). This highlights the importance of the adaptive protocol in the presence of partially correlated noise. Also, we see that in both cases, $\Delta r$ vanishes at $\mu=1$, confirming that there is no gain in the fully correlated scenario.

The similar benefits can also be reported for the LM05 protocol in the CBPF case -- the adaptive scheme enhances the secret key rate as well as the critical value of $p$.

$(3)$ \textit{$p$-dependence:} Similar to the SDC scheme, the key rate decreases with increasing $p$ and vanishes after a certain $p_{cr}$ for the CBPF noise. The adaptive scheme improves both the key rate and $p_{cr}$ values, for all values of $\mu$. We observe that the $p_{cr}$ value increases with $\mu$ for the adaptive protocol, as evident from Fig. \ref{fig:partial-correlated}(e). For instance, $p_{cr} \,(\mu=0.7) > p_{cr} \,(\mu=0.35)$. We find the peak values of $\Delta r$ at the $p$ values where $r_{conventional}=0$, shortly after which it vanishes (i.e., $r_{adaptive}$ also vanishes). 

$(4)$ \textit{$\mu$-dependence:} For low values of $p$ (e.g. $0<p<0.13$), we observe an advantage in the key rate with the adaptive scheme over the whole range of $\mu$. With an increase in $p$, there exists a certain threshold value of $\mu$, below which both the adaptive and conventional key rates vanish. However, $r_{adaptive}$ becomes positive for a lesser value of $\mu$ compared to $r_{conventional}$, implying that lesser correlations between channels are enough to obtain a positive key rate via the adaptive method. For example, for $p=0.15$, in the region $0.1<\mu<0.7$, $r_{adaptive}>0$, while $r_{convetional}=0$. Hence, with the adaptive scheme, secure transmission is achieved for lesser degrees of correlation in the channel, emphasizing its importance in the presence of correlated noise channels.

\section{Action of Non-Markovianity in NAQKD}\label{sec:nonMarkovian}

When quantum systems are affected by the environment, depending on whether the transmission channels retain memory of the past system dynamics, noise models are often categorized as Markovian~\cite{Nielsen_2010_book}, which exhibit no memory effects, or Non-Markovian~\cite{theory_of_open_quantum_10.1093/acprof:oso/9780199213900.001.0001} (NM), which retain the memory of earlier stages of evolution, thereby influencing subsequent noise processes. Unlike Markovian processes, in which information flows irreversibly from the system to the environment, NM dynamics permits temporary backflow of information, leading to correlations between consecutive channel uses. In several quantum information tasks, non-Markovianity has been shown to retain quantum features in the system, compared to the Markovian regime~\cite{abhishek_muhuri_PhysRevA.109.032616}.

We study here the effectiveness of the noise-adaptive protocol in the presence of local non-Markovian noise models
in terms of the secret key rate. For this, we can consider the Kraus operator decomposition of the NM phase flip channel given by~\cite{shrikant_PhysRevA.98.032328}
\begin{equation}\label{eq: non markovian}
      \hspace{-1.5em}  \hat{K}_I(t) = \sqrt{(1-\alpha p)(1-p)}\,\mathbb{I};\;
        \hat{K}_Z(t)=\sqrt{\left(1+\alpha(1-p)\right) p}\,\hat{\sigma}^Z
\end{equation}
where $0\leq \alpha \leq 1$ represents the degree of Non-Markovianity, and, $p$ the time dependent noise parameter. We can readily see that for $\alpha \to 0$, Eq. \eqref{eq: non markovian} reduces to the conventional phase flip noise action. The NM bit flip and bit-phase flip channels can be defined in a similar way, by replacing $\hat{\sigma}^Z$ with $\hat{\sigma}^X$ and $\hat{\sigma}^Y$ in Eq.\eqref{eq: non markovian}, respectively.
We find that the advantage in the secret key rate obtained in the case of independent noise models across the three protocols (SDC, LM05, and 2-way BB84) persists even after the introduction of non-Markovianity in the channels. 
Further, no advantage of the adaptive scheme in the case of NM bit flip noise continues to hold.
In the case of the SDC protocol, under NM phase flip noise, as $\alpha$ increases, $\Delta r$ approaches the peak value earlier (i.e., for a lower value of $p$), and subsequently drops to zero more rapidly. However, it again rises and at maximum noise strength $p=0.5$, $\Delta r$ is positive for $\alpha=1$, followed by the corresponding values in the decreasing order of $\alpha$, as depicted in Fig. \ref{fig:nonmark-all}(a). The maximum value of $\Delta r$ is not affected by non-Markovianity, i.e., it stays the same for all values of $\alpha$.

The main findings in the case of NM bit-phase flip noise are similar, although there are a few differences across the three protocols. In case of all three protocols, the peak value of $\Delta r$ is attained at lower values of $p$, with increasing $\alpha$, as earlier. However, for the SDC and the LM05 protocols, the limiting noise strength for secure transmission reduces with increasing $\alpha$. As we can see from the Figs. \ref{fig:nonmark-all} (b) and (c), with increasing $\alpha$, $\Delta r$ attains the peak and drops to zero much faster. Once it reaches zero, there is no further rise, in contrast to what we observed for NM phase flip noise in the SDC protocol. Thus, for these two cases, the secure transmission noise limit reduces with amplifying non-Marovianity.

The observations are different in the case of the 2-way BB84 protocol. In this case, after collapse (i.e., vanishing $\Delta r$), following the peak, the $\Delta r$ value revives for $\alpha >0$, similar to the SDC protocol under NM phase flip noise (see Fig. \ref{fig:nonmark-all}(d)). At $p=0.5$, $\Delta r$ is non-vanishing for higher values of $\alpha$. As discussed in Sec. \ref{sec: independent improvement}, the gain in the key rate from the adaptive BB84 protocol is the highest among all, and this observation continues to hold in the non-Markovian scenario as well.

In all of these cases, the optimal unitaries are the same as found for the corresponding independent channel cases and listed above in Table \ref{table:unitaries}. Although there is no physical improvement in $\Delta r$ or $p_{max}$ (for secure transmission) due to non-Markovianity, for the SDC protocol under NM phase flip, and for the 2-way BB84 protocol under the NM bit-phase flip noise, at $p=0.5$ (optimal noise strength), a greater value of adaptive key rate is obtained for a higher $\alpha$.

For low noise strength, the sharp increase of $\Delta r$, indicating the beneficial role of the adaptive scheme, with increasing non-Markovian strength, can be regarded as the signature of non-Markovianity in QKD.

\begin{figure*}[t]
    \centering
    \includegraphics[width=0.8\linewidth]{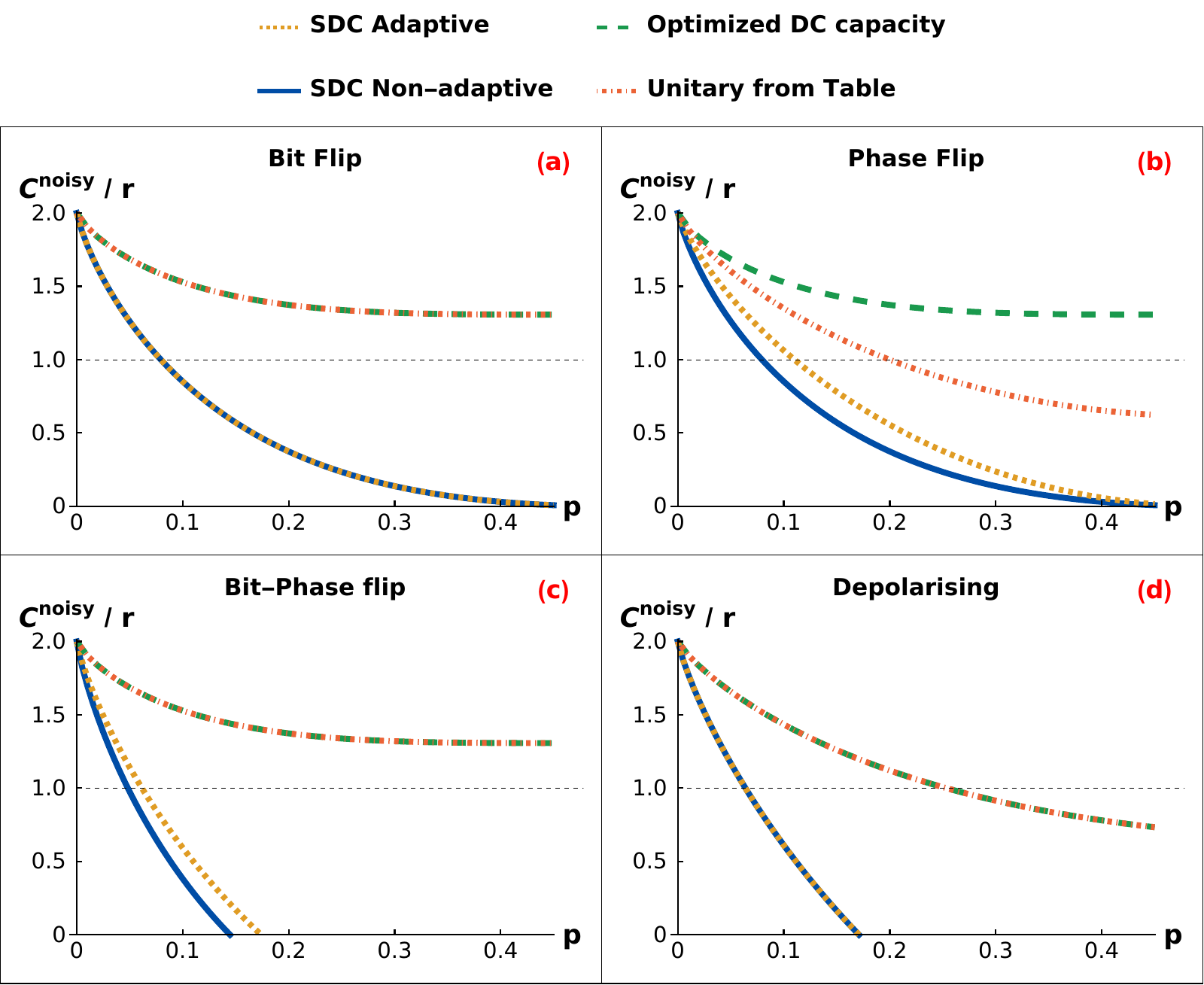}
    \caption{\justifying{(Color online) 
    Comparison of the lower bounds on the key rates for the noise-adaptive and conventional non-adaptive protocols, as well as the noisy dense coding capacity. The quantities $r_{\mathrm{adaptive}}$ (orange dotted line), $r_{\mathrm{conventional}}$ (blue solid line), and $C^{\mathrm{noisy}}$ (green dashed line) are plotted as functions of the noise parameter $p$ for (a) bit flip, (b) phase flip, (c) bit-phase flip, and (d) depolarizing channels. In all cases, $C^{\mathrm{noisy}}$ significantly exceeds both the adaptive and non-adaptive key rates, while the black dotted line denotes the classical limit of dense coding capacity. These plots provide a qualitative comparison of the extent to which classical information transmission can be converted into secure information transmission. The red dot-dashed line represents an analogous quantity to the noisy dense coding capacity, where, without obtaining the $U_B^{min}$, we choose unitaries from Table~\ref{table:unitaries}. These results also indicate that adaptive schemes have to be developed depending on the quantum protocols, since the unitaries that optimise the key rate may not be optimal for classical information transmission without security. }}
    \label{fig:SDC-comparison}
\end{figure*}
\section{Comparison of optimal unitaries of NAQKD with noisy dense coding capacity}
\label{sec:DCcapacity-Comp}

Let us find out whether the noise adaptive key rate $r_{adaptive}$ for secure dense coding protocol, involving a single sender Bob ($B$) to a single receiver Alice ($A$), is connected with the noisy quantum super dense coding capacity, 
~\cite{Shadman2010,Shadman2011,Shadman2012,Shadman2013, Das_PRA_2014_noise-inverts-dense-coding,Das_distributed-PhysRevA.92.052330}. 

For a shared state $\rho_{AB} =\Lambda^f_{A' \rightarrow B} (\rho_{AA'})$, where $\rho_{AA'} = |\phi^+\rangle \langle\phi^+|_{AA'}$ is the maximally entangled state, 
where $\Lambda^f_{A' \rightarrow B}$ is the forward transmission channel, and the presence of noise actually transforms the resource Bell state to $\rho_{AB}$, finally shared between Alice and Bob. Now Bob performs unitary encoding to encode his binary two-bit messages (in each run) he wants to send to Alice without considering the security aspects of it, i.e., without performing any security check run, and classical post-processing procedure. Again, while transmitting his encoded state back to Alice, it can be affected by the backward channel $\Lambda^b_{B \rightarrow A}$. Alice performs measurements to access the maximal amount of information, and that maximal information which can be sent in the presence of noise is known as the noisy dense coding capacity, given by
\begin{equation}
C^{\mathrm{noisy}}(\rho_{AB}) =  \log_2 d_{B} + S(\rho_A) - S(\tilde{\rho}_{AB}),
\label{eq:noisyDC}
\end{equation}
where
\begin{equation}
\tilde{\rho} = \Lambda^b \Big( (I_A \otimes U^{\min}_{B})\, \rho_{AB}\, (I_A \otimes U^{\min }_{B})^\dagger \Big).
\end{equation}
Here $\Lambda^b$ denotes the noisy backward quantum channel acting on Bob's subsystem, and $U^{\min }_{B}$ denotes the unitary operator on the
sender's (Bob) side, which minimizes the von Neumann entropy of $\Lambda^b \Big( (I_A \otimes U_{B})\, \rho_{AB}\, (I_A \otimes U_{B})^\dagger \Big)$.

Comparing $C^{\mathrm{noisy}}$ and $r$, we observe that for all noise models considered, 
$C^{\mathrm{noisy}}$ 
(green dashed line in Fig. \ref{fig:SDC-comparison}) significantly exceeds both the adaptive (orange dotted line in Fig. \ref{fig:SDC-comparison}) and non-adaptive (blue solid line in Fig. \ref{fig:SDC-comparison}) key rates. 
This is due to the fact that $C^{\mathrm{noisy}}$ quantifies the total classical information that can be transmitted using the shared resource and optimal encoding; it does not account for information leakage to the environment due to the noise in the transmission channel. This information can be accessed by any malicious eavesdropper, and she can, in principle, have the knowledge of the messages shared between Alice and Bob. 
In contrast, the key rate of a secure dense coding protocol is further constrained by secrecy requirements and is typically bounded by a Devetak--Winter \cite{Devetak_PRSA_2005_key-rate, Beaudry_PRA_2013_two-way-QKD} type expression, given in Eq. \eqref{eq:devetak-key}, where the first quantity 
denotes the mutual information between Alice and Bob, whereas the second quantity (can also be expressed as $\chi(B:E)$; the Holevo quantity) bounds the information accessible to the eavesdropper.

As a consequence, the secure key rate, even the improved one $r_{adaptive}$, is generally upper bounded by the $C^{\mathrm{noisy}}$. The gap between these two quantities increases with the increase of noise parameters, which indicates the fact that the correlations between the honest parties decrease, and any additional third party (or environment) can share some amount of correlations, through the noisy channels $\Lambda$, as reflected by the entropy term $S(\tilde{\rho})$. 
The plots in Fig. \ref{fig:SDC-comparison} distinguish the portion of the dense coding capacity that contributes to the secure communication  with respect to the increasing noise parameters, which we have shown to be improved due to the optimal choice of encoding and decoding operation. 

It is interesting to investigate whether the adaptive unitary optimizing Eq. (\ref{eq:working-key-rate-maximized}) is the same as the optimal unitary that maximizes Eq. \eqref{eq:noisyDC}. In order to find the answer for all noise models, we compute $C^{\mathrm{noisy}}(\rho_{AB})$ by replacing
the unitary $U_{B}^{min}$ in the third term of Eq. \eqref{eq:noisyDC} by the optimal unitary $W(\theta, \chi,\phi)$ as specified in Table \ref{table:unitaries} (as shown in Fig. \ref{fig:SDC-comparison} by red dot-dashed line). 


For the bit flip and the depolarizing channels, the key rates do not show any improvement, which means that $W = \mathbb{I}_2$. Similarly, we find that among the several $U_{B}^{min}$, that give rise to the same $C^{\mathrm{noisy}}$, $\mathbb{I}_2$, is also one of them (as it is visible by comparing the green dashed and the red dot-dashed lines in Fig. \ref{fig:SDC-comparison}(a) and (d)).
On the other hand, the noise adaptive key rate is found to be better than the non-adaptive one for both the phase flip and bit-phase flip channels. Interestingly, we observe that although  $U_{B}^{min}$ does not optimize the key rate in case of phase flip channel (see Fig. \ref{fig:SDC-comparison}(b)), all the adaptive unitaries given in Table \ref{table:unitaries}, for bit-phase flip channel, turns out to be the optimal unitary $U_{B}^{min}$, as evident from Fig. \ref{fig:SDC-comparison}(c).
Notice further that $C^{\mathrm{noisy}} > 1$ indicates the quantum advantage (the black dotted line 
represents the classical capacity in Fig. \ref{fig:SDC-comparison}  ). The noisy dense coding protocol is advantageous over the entire range of $p$, for bit flip, phase flip, and bit-phase flip noise, whereas for the depolarizing channel, it is not advantageous for $p > 0.255$.

\section{Conclusion}
\label{sec:conclusion}

Depending on the nature of the noise, adapting the encoding and decoding procedures can significantly improve the performance of several quantum-information-processing tasks like dense coding~\cite{Shadman2010,Shadman2011,Shadman2012, Das_PRA_2014_noise-inverts-dense-coding,Das_distributed-PhysRevA.92.052330}, quantum error correction ~\cite{Mandayam-noise-PhysRevResearch.6.043034, Jayashankar2022NoiseAdaptedQEC} compared to protocols employing fixed strategies. Motivated by this observation, we developed a general framework for noise-adaptive two-way deterministic quantum key distribution (NAQKD) protocols against collective eavesdropping attacks. In particular, the honest parties are allowed to optimize the encoding and decoding operations according to the characteristics of the noisy quantum channels to maximize the secure key rate. We investigated this for three representative two-way QKD protocols, secure dense coding (SDC), LM05, and two-way BB84.


 Our investigations reveal that the usefulness of adaptive protocols strongly depends on the structure of the noise affecting the forward and backward transmission channels. For independent and identical noise acting on both honest parties' channels, we identified classes of Pauli channels for which adaptive optimization offers no advantage across all considered protocols, whereas phase flip and bit-phase flip channels exhibit clear noise-dependent improvements. Moreover, non-unital noise models, particularly amplitude-damping channels, lead to substantial enhancement of the secure key rates in LM05 and two-way BB84 protocols.


We further extended our analysis to correlated noisy channels. While fully correlated channels do not exhibit any adaptive advantage, partially correlated channels were shown to possess parameter regimes where the adaptive protocols outperform their non-adaptive counterparts, especially for the SDC and LM05 schemes. 
In addition, by incorporating non-Markovian dynamics, we identified regimes in which memory effects in the environment can further enhance the performance of the adaptive QKD protocol over the conventional ones. In all these cases, we determined the classes of optimal encoding unitaries responsible for the observed adaptive advantage.

 It is important to emphasize that, in two-way protocols, an enhancement in the key rate through the adaptive method can only be observed when the transmission channel is affected by noise. Since the actual information is carried by the encoded states transmitted through the backward channel, the effect of noise and the implementation of adaptive strategies become particularly important for the backward transmission. In contrast, as no encoding is performed in the forward channel, eavesdropping attacks acting solely on it do not reveal useful information to the adversary, apart from reducing the key rate. 

Overall, our results demonstrate that noise-adaptive encoding and decoding operations can significantly enhance the secure key rate across a broad range of realistic noise models and two-way QKD protocols. The enhancement reported in secure key generation highlights the practical relevance of adaptive strategies for future long-distance and experimentally realizable scalable quantum communication networks.

\section*{Acknowledgment}


 A.P. acknowledge partial support from the ``INFOSYS Scholarship for senior students''. A.P. and A.S.D. acknowledge support from the project entitled "Technology Vertical - Quantum Communication'' under the National Quantum Mission of the Department of Science and Technology (DST)  (Sanction Order No. DST/QTC/NQM/QComm/2024/2 (G)).

\bibliographystyle{apsrev4-1}
\bibliography{ref}

\newpage

\appendix

\begin{widetext}
\section{Generalized purification scheme for encoding operation}\label{app:generalized-purification}
In this section, we are going to prove Eq. \eqref{eq:purification-general}. Suppose Bob receives the subsystem $B$ of the shared state $\rho_{AB}$ coming from Alice via channel $\Lambda^f$ for the encoding process. Assume that the state $\rho_{AX}$ is the same as $\rho_{AB}$, with the index $B$ being replaced by $X$.
In the non-purified scheme, Bob applies the generalized encoding operator $U^{xy}$ on it. However, in the purified version, Bob will perform a measurement-based encoding scheme after concatenating the incoming state with an auxiliary state $\rho_{\phi^{+}}$, the proof of which is shown below. We can write $\rho_{AX} = \sum_{ij} |i\rangle\langle i|_X\, \rho_{AX}\, |j\rangle \langle j|_X = \sum_{ij} \rho_A^{ij}|i\rangle\langle j|_X$, and $|\phi^+\rangle_{X'B}=\frac{1}{\sqrt{2}}\sum_p |pp\rangle_{X'B}$. Then we want to prove that
\begin{equation}\label{}
    4 \times {}_{XX'}\langle \phi(xy)|\left(\rho_{AX} \otimes |\phi^{+}\rangle\langle \phi^{+}|_{X'B}\right)|\phi(xy)\rangle_{XX'} \sim (\mathbb{I}_{A}\otimes U_B^{xy} )\rho_{AB} (\mathbb{I}_{A}\otimes U_B^{xy\,\dagger} ) \equiv \rho_{AB}^{xy}.
\end{equation}
Now from Eq. \eqref{eq:Bellwhat}, writing $|\phi(xy)\rangle_{XX'}=U_X^{xy\,\dagger}\otimes \mathbb{I}_{X'} |\phi^+\rangle_{XX'}$, we obtain
\begin{eqnarray}
    &&{}_{XX'}\langle \phi^+| U^{xy}_X\otimes \mathbb{I}_{X'}\, (\rho_{AX} \otimes |\phi^+\rangle\langle\phi^+|_{X'B}) U_X^{xy\,\dagger}\otimes \mathbb{I}_{X'} |\phi^+\rangle_{XX'}\\
    &=&{}_{XX'}\langle \phi^+| U^{xy}_X\otimes \mathbb{I}_{X'}\, (\sum_{ij} \rho_A^{ij}|i\rangle\langle j|_X\, \otimes\,\frac{1}{2}\sum_{pp'}|p\rangle\langle p'|_{X'})\,U_X^{xy\,\dagger}\otimes \mathbb{I}_{X'} |\phi^+\rangle_{XX'}\: |p\rangle\langle p'|_{B} \\
    &=& \sum_{\substack{i,l,l',\\j,p,p'}} \frac{1}{2}\, {}_{XX'}\langle ll|\,U^{xy}_X\otimes \mathbb{I}_{X'} \,|ip\rangle\langle jp'|_{XX'}\, U_X^{xy\,\dagger}\otimes \mathbb{I}_{X'}\,|l'l'\rangle_{XX'} \,\left[\frac{1}{2}\rho_A^{ij} |p\rangle\langle p'|_{B}\right]\\
    &=& \sum_{ijpp'} \left( \frac{1}{2}\sum_{ll'}\, \langle l|U^{xy}_X|i\rangle\,  \langle j |U_X^{xy\,\dagger}|l' \rangle\, \delta_{lp}\, \delta_{p'l'} \right)\left[\frac{1}{2}\rho_A^{ij} |p\rangle\langle p'|_{B}\right]\\ 
    &=& \frac{1}{4}\sum_{ijpp'} \rho_A^{ij} \left( \langle p'| U^{xy}|i\rangle\langle j |U^{xy\,\dagger}|p'\rangle \right) |p\rangle\langle p'|_B \\
    &=& \frac{1}{4}\sum_{ijpp'} \rho_A^{ij} |p\rangle\langle p'| U^{xy}_B|i\rangle\langle j |U^{xy\,\dagger}_B|p'\rangle \langle p'| = \frac{1}{4} (\mathbb{I}_{A}\otimes U_B^{xy} )\rho_{AB} (\mathbb{I}_{A}\otimes U_B^{xy\,\dagger}).
\end{eqnarray}

\section{Proof of the completeness condition of the rotated Bell basis}\label{app:completeness-orthounitary}
The unitarily rotated Bell states can be written as $|\phi(xy)\rangle= \frac{1}{\sqrt{2}} \sum_{l=0}^{1} U^{xy\,\dagger} |l\rangle \otimes |l\rangle$. If the $\{U^{xy}\}_{x,y = 0}^1$, are a complete set of orthogonal unitary operators, then 
\begin{eqnarray}
    \sum_{x,y = 0}^1 |\phi(xy)\rangle \langle\phi(xy)| &=& \sum_{x,y = 0}^1 \left(\frac{1}{\sqrt{2}} \sum_{l=0}^{1} U^{xy\,\dagger} |l\rangle \otimes |l\rangle \right)\left(\frac{1}{\sqrt{2}} \sum_{l'=0}^{1}  \langle l'|U^{xy} \otimes \langle l'| \right) \label{cmp2}\\
    &=& \sum_{l,l'=0}^{1} \left(\frac{1}{2} \sum_{x,y = 0}^1 U^{xy\,\dagger} |l\rangle\langle l'|U^{xy} \otimes |l\rangle \langle l'| \right) \label{cmp3}\\
    &=& \sum_{l,l'=0}^{1} \delta_{ll'}  I_2  \otimes |l\rangle \langle l'| = I_4,
\end{eqnarray}
where from Eqs. \eqref{cmp2} to \eqref{cmp3}, we have used the completeness relation of the orthogonal unitary.

\section{Proof of $p(ij|xy)= \delta_{ix}\delta_{jy}$}\label{app:proofdelta}

From Eq. \eqref{eq:BellAlice}, it is clear that $|\chi(ij)\rangle_{AA'} = \mathbb{I}_A \otimes U^{ij}_{A'}|\phi^+\rangle_{AA'} $. Now in the noiseless scenario ($\Lambda^f_{A'\rightarrow B}=\mathbb{I}=\Lambda^b_{B\rightarrow A'}$), the final encoded state received by Alice is given by
\begin{equation}
    \rho_{AA'}^{xy}=(\mathbf{I}_A \otimes U_{A'}^{xy})\,|\phi^+\rangle\langle\phi^+|_{AA'}\,(\mathbb{I}_A\otimes U_{A'}^{xy\,\dagger}).
\end{equation}
Therefore, the measurement probabilities at Alice's end read as
\begin{eqnarray}
    p(ij|xy) &=&_{AA'}\langle\chi(ij)| \rho_{AA'}^{xy}|\chi(ij)\rangle_{AA'}\\
    &=& \left| _{AA'}\langle\chi(ij)|\,\mathbb{I}_A \otimes U_{A'}^{xy}\,|\phi^+\rangle_{AA'}  \right |^2\\
    &=& \left|_{AA'}\langle\phi^+|\, \mathbb{I}_A \otimes U^{ij\,\dagger}_{A'} U^{xy}_{A'} |\phi^+\rangle_{AA'} \right |^2\\
    &=& \left| \text{tr}_{AA'} \left(\mathbb{I}_A \otimes U^{ij\,\dagger}_{A'}\, U^{xy}_{A'} \,\otimes |\phi^+\rangle\langle\phi^+|_{AA'} \, \right) \right|^2\\
    &=&\left|\frac{1}{2} \text{tr}_{A'} (U^{ij\,\dagger}_{A'}\, U^{xy}_{A'}) \right |^2
    = \delta_{ix}\delta_{jy},
\end{eqnarray}
where from the fourth to the fifth equality, we have used the fact that $\text{tr}_{A} (|\phi^+\rangle\langle\phi^+|_{AA'}) = \frac{1}{2}\mathbb{I}_{A'}$, and in the last equality, we have used the orthogonality condition. 

\section{Proof of  Eq. \eqref{eq:orthogonal_purification}}\label{app:proof_ortho_puri}
From Eq. \eqref{eq:purification-general}, we have 
\begin{eqnarray}\label{eq:puri}
    \rho_{AB}^{xy} &=& {}_{XX'}\langle \phi(xy)|\left(\rho_{AX} \otimes |\phi^{+}\rangle\langle \phi^{+}|_{X'B}\right)|\phi(xy)\rangle_{XX'} \nonumber \\
    &=& {}_{XX'}\langle B(xy)|\rho_{AX} \otimes (\mathbb{I}_{X'}\otimes W_B)\,|\phi^{+}\rangle\langle \phi^{+}|_{X'B}\,(\mathbb{I}_{X'}\otimes W_B^{\dagger} ) |B(xy)\rangle_{XX'}. 
\end{eqnarray}
We know $|\phi(xy)\rangle_{XX'}=U_X^{xy\,\dagger}\otimes \mathbb{I}_{X'} |\phi^+\rangle_{XX'}$. Putting this above in the LHS, we have
\begin{eqnarray}
    &&{}_{XX'}\langle\phi^+| U_X^{xy}\otimes \mathbb{I}_{X'}  \left(\rho_{AX} \otimes |\phi^{+}\rangle\langle \phi^{+}|_{X'B}\right)\, U_X^{xy\,\dagger}\otimes \mathbb{I}_{X'} |\phi^+\rangle_{XX'}\\
    &=&{}_{XX'}\langle \phi^+| \mathbb{I}_X\otimes U_{X'}^{xy\,T}\left(\rho_{AX} \otimes |\phi^{+}\rangle\langle \phi^{+}|_{X'B}\right)\,\mathbb{I}_X\otimes U^{xy\,\ast}_{X'}|\phi^+\rangle_{XX'}\\
    &=&{}_{XX'}\langle \phi^+| \mathbb{I}_X\otimes (W\hat{\sigma}^{xy})_{X'}^{T}\left(\rho_{AX} \otimes |\phi^{+}\rangle\langle \phi^{+}|_{X'B}\right)\,\mathbb{I}_X\otimes (W\hat{\sigma}^{xy})^{\ast}_{X'}|\phi^+\rangle_{XX'}\\
    &=&{}_{XX'}\langle B(xy)| \left[\rho_{AX} \otimes\, (W^T_{X'}\otimes \mathbb{I}_B) \,|\phi^{+}\rangle\langle \phi^{+}|_{X'B}\, (W^{\ast}_{X'}\otimes \mathbb{I}_B)\, \right]\,|B(xy)\rangle_{XX'}\\
    &=&{}_{XX'}\langle B(xy)| \left[\rho_{AX} \otimes\, (\mathbb{I}_{X'}\otimes W_B) \,|\phi^{+}\rangle\langle \phi^{+}|_{X'B}\, (\mathbb{I}_{X'}\otimes W^{\dagger}_B)\, \right]\,|B(xy)\rangle_{XX'},
\end{eqnarray}
where $U^{xy}=W\hat{\sigma}^{xy}$, and we have used the invariance property of the Bell state $|\phi^+\rangle$ (for a unitary operator $\hat{V}$); 
\begin{eqnarray}
    && (\hat{V}\otimes\mathbb{I})|\phi^+\rangle = (\mathbb{I}\otimes \hat{V}^T)|\phi^+\rangle\\
    &\longleftrightarrow& \langle\phi^+| (\hat{V}^\dagger \otimes \mathbb{I}) = \langle\phi^+|(\mathbb{I}\otimes \hat{V}^{\ast})\,,
\end{eqnarray}
and, the fact that $|B(xy)\rangle = (\mathbb{I}\otimes \hat{\sigma}^{\ast\,xy})|\phi^+\rangle = (\mathbb{I}\otimes \hat{\sigma}^{xy})|\phi^+\rangle$ (as $\hat{\sigma}^{xy} ~ \forall ~x,y \in \{0,1\}$, are chosen in such a way that $\hat{\sigma}^{xy\,\ast} = \hat{\sigma}^{xy}$). Here, $\hat{\sigma}^{xy}$ are the Pauli matrices along with identity, and $|B(xy)\rangle$ represents the Bell states. 

\section{Purification of test run}\label{pur_test_puri}

We can develop the purified version of the noise adaptive test run or the security check run, similar to Appendix \ref{app:proof_ortho_puri}. 
It is important to note that Bob measures in the eigen basis of $\hat{\sigma}^Z$, on the state he receives from Alice, and then prepare another state in the eigenbasis of $W(\theta,\chi,\phi)\hat{\sigma}^X W(\theta,\chi,\phi)^\dagger$, and sends to Alice through the backward quantum channel $\Lambda^{b}_{B\rightarrow A'}$. Note that the eigenbases of $W(\theta,\chi,\phi)\hat{\sigma}^X W(\theta,\chi,\phi)^\dagger$ are denoted as $W(\theta,\chi,\phi)|y_\vdash\rangle $, for $ y = 0,1 $ where $|0_\vdash \rangle = |+\rangle$, and $|1_\vdash \rangle = |-\rangle$.

Measurement by Bob, first in the eigenbasis of $\hat{\sigma}^Z$, in his part of the shared state, and prepare it in the $W(\theta,\chi,\phi)|j_\vdash\rangle $, for $j = 0,1$, can be written as 
\begin{eqnarray}
    {}_X\langle x|\rho_{AX} |x \rangle_X \otimes W(\theta,\chi,\phi)|y_\vdash\rangle \langle y_\vdash| W(\theta,\chi,\phi)^\dagger_B.\label{eq:test-prepare}
\end{eqnarray}
Note that here, the measured state, although performed by Bob, we have used the subscript $X$, and the prepared state we put the subscript $B$. Also, it is worth mentioning the fact that the Bell state $|\phi^+\rangle$ is $U \otimes U^*$ invariant, where $U$ is any SU(2) matrix, and $U^*$ is the complex conjugation of $U$. Hence preparing the state in $W(\theta,\chi,\phi)|y_\vdash\rangle$, the basis is equivalent to measuring one subsystem of $|\phi^+\rangle$, in the $W(\theta,\chi,\phi)^*|y_\vdash\rangle$ basis. Hence, Eq. \eqref{eq:test-prepare}, can be written as 
\begin{eqnarray}
   && {}_X\langle x|\rho_{AX} |x \rangle_X \otimes W(\theta,\chi,\phi)|y_\vdash\rangle \langle y_\vdash| W(\theta,\chi,\phi)^\dagger_B \nonumber \\
    &=& {}_X\langle x|\rho_{AX} |x \rangle_X \otimes {}_{X'}\langle y_\vdash |W^T_{X'} \otimes \mathbb{I}_B|\phi^+\rangle \langle\phi^+|_{X'B} W^*_{X'} \otimes \mathbb{I}_B |y_\vdash\rangle_{X'} \\
    &=& {}_{XX'}\langle x, y_\vdash|\left[\rho_{AX}  \otimes (W^T_{X'} \otimes \mathbb{I}_B) |\phi^+\rangle \langle\phi^+|_{X'B} (W^*_{X'} \otimes \mathbb{I}_B) \right]|x,y_\vdash\rangle_{XX'} \\
    &=& {}_{XX'}\langle x, y_\vdash|\left[\rho_{AX} \otimes\, (\mathbb{I}_{X'}\otimes W_B) \,|\phi^{+}\rangle\langle \phi^{+}|_{X'B}\, (\mathbb{I}_{X'}\otimes W^{\dagger}_B)\, \right]|x,y_\vdash\rangle_{XX'}.
\end{eqnarray}

Hence we can conclude that, similar to the key generation run, at the test run, the incoming state $\rho_{AX}$, first concatenated with the unitary rotated auxiliary state  $\rho^{\phi^+}_{X'B}$, by $W^*_{B}$, in part $B$, and measured in the common parties $XX'$, by the measurement basis $|x,y_\vdash\rangle$, where $x,y \in \{0,1\}$.
The resultant state has now been sent back to  Alice via a backward quantum channel $\Lambda^b_{B\rightarrow A'}$. Therefore, the final measurement by Alice is performed by using the measurement operator $\,\hat{\tilde{\mathcal{G}}}_{AA'}$, given in Eq.\eqref{eq:GBob}, to incorporate the effect of noise arising from the quantum channels.

\section{Dense coding based noise adaptive quantum key distribution protocol}\label{app:SDC}

The quantum super dense coding-based secure QKD protocol has been initially developed in \cite{Beaudry_PRA_2013_two-way-QKD} (The Paper~\cite{PhysRevLett.89.187902} first introduced the idea of Deterministic Secure Direct Communication Using Entanglement.). 
For a given noise model, $\Lambda$, with noise elements $\{E_i\}$, affecting the forward and backward transmission channels involved in the two-way physical transmission of a qubit, the noise-adaptive secure dense coding protocol proceeds as follows:

\begin{enumerate}
    \item[(a).] \textbf{Preparation:} Similar to quantum dense coding  \cite{Bennett_PRL_1992_dense-coding}, Alice prepares the maximally entangled Bell state, $|\phi^+_{AA'}\rangle = \frac{1}{\sqrt{2}}(|00\rangle + |11\rangle)_{AA'}$, keeps one qubit ($A$) in her quantum memory and sends the other ($A'$) to Bob via  $\Lambda^f_{A' \rightarrow B}$, resulting in shared $\rho_{AB} =\Lambda^f_{A' \rightarrow B} (\rho_{AA'})$, where $\rho_{AA'} = |\phi^+_{AA'}\rangle \langle\phi^+_{AA'}|$.

    \item[(b).] Upon receiving the subsystem $B$ of $\rho_{AB}$,  Bob performs the key generation runs almost every time with probability $c \approx 1$, and a few times with probability $1 - c$, the test or security check run. \\
    In the \textbf{key generation run}, he performs
    any one of the unitary operators, $U^{xy}= W(\theta,\chi,\phi)\cdot \hat{\sigma}^{xy}$, with equal probability to encode the classical raw key pair $(x,y)$, where $x,y\in\{0,1\}$, and sends it back to Alice through the backward transmission channel $\Lambda^{b}_{B\rightarrow A'}$. Here $\hat{\sigma}^{00}=\mathbb{I},\, \hat{\sigma}^{01}=\hat{\sigma}^Z,\,\hat{\sigma}^{10}=\hat{\sigma}^X,\, \hat{\sigma}^{11}=-i \hat{\sigma}^Y$. In this step Bob's unitary operation leads to a shared ensemble of states, $\{p^{xy} = \frac 14,\rho_{AB}^{xy}\}$, where, $\rho_{AB}^{xy}=(\mathbb{I}_A\,\otimes U^{xy}_B)\,\rho_{AB}\,(\mathbb{I}_A\,\otimes U^{xy\,\dagger}_B)$, are mutually orthogonal encoded states in the noiseless scenario.
    
    In the \textbf{test run}, Bob performs a projective measurement on his qubit in the $\hat{\sigma}^Z$ basis, and prepares an eigenstate of $W(\theta,\chi,\phi)\hat{\sigma}^X W(\theta,\chi,\phi)^\dagger$, and sends it to Alice via $\Lambda^{b}_{B\rightarrow A'}$.

    Note that in the usual secure DC based protocol, Bob encodes the raw key bit using only the Pauli matrices and the identity operator, irrespective of whether the transmission channel is noiseless or noisy.  To extract the maximal amount of secure key, even in the presence of noise, we have introduced a general $SU(2)$ unitary $W(\theta,\chi,\phi)$, termed as the \emph{adaptive unitary}. We will optimize Eq.~\eqref{eq:working-key-rate} over the values of $\theta,\chi,\text{and }\phi$, in order to obtain the highest amount of key rate, for various kinds of noise models. 
    
    \item[(c).] Alice receives the qubit sent by Bob through the quantum channel $\Lambda^b_{B\rightarrow A'}$, thereby resulting in the ensemble $\{p^{xy},\rho_{AA'}^{xy}\}$ (key generation run). This step can be represented as $\Lambda^b_{B \rightarrow A'} (\rho_{AB}^{xy}) = \rho_{AA'}^{xy}$.
    
    \item[(d).] Alice again chooses randomly two operations, one is for the key generation run with very high probability, and the other one is for the test run with the remaining $1 - c$ probability. She
    performs rotated Bell basis measurements, $\mathcal{M}_{AA'} = \{M_{AA'}^{ij}\}$, given in Eq. \eqref{eq:BellAlice}, which can also take the form of $|\chi(ij)\rangle  = I\otimes W(\theta,\chi,\phi) |B(ij)\rangle$.

     For the test run, Alice measures her stored qubit in the $\hat{\sigma}^Z$-basis, and the received qubit in the rotated $W(\theta,\chi,\phi)\hat{\sigma}^X W(\theta,\chi,\phi)^\dagger $ basis. The optimal choice of unitary  $W(\theta,\chi,\phi)$, definitely depends on the noise model, and the noise strength; but in the noise adaptive protocol, we will assume that Alice and bob already agreed with a particular values of  $W(\theta,\chi,\phi)$, by predicting the possible noise model in their transmission channel. 
\end{enumerate}
It is worth mentioning the fact that the honest parties choose to perform the security check run or test run to estimate the presence of noise in the transmission channel, or any possible attack by the Eavesdropper.\\ 
An eavesdropper, Eve, can attack the quantum transmission channel in two different ways: by tampering with the traveling qubit before the encoding operation, to alter the shared state, or by measuring the qubit after encoding to extract information about the message. The eavesdropping process is possible using any strategy allowed by quantum mechanics, and here, in the worst-case scenario, we will consider that Eve can control all the possible interfaces of the purification of the shared state between Alice and Bob. 

\textbf{Post-processing and key rate:} After a sufficient amount of quantum key distribution protocol, involving both the key generation run and the test runs performed by the honest parties randomly, any one of the honest parties starts a one-way classical post-processing protocol, which involves shifting of keys, classical error corrections, and privacy amplification. In this process, the honest parties will estimate the possible lower bound on the secret key rate of their protocol, according to the Devetak-Winter formula~\cite{Devetak_PRSA_2005_key-rate}, which will then be optimized over the parameters $\theta$, $\chi$, $\phi$, of  $W(\theta,\chi,\phi)$,  for different noise models. 
If the key rate turns out to be positive, they will proceed further with the post-processing events; otherwise, they will abort the protocol. 

In the following, we will calculate the lower bound on the secure key rate based on the purified version of the noise-adaptive SDC protocol, which is one of the special cases of the generalized two-way noise adaptive protocol described in Sec. \ref{Sec:Generic-noise-adaptive-protocol}.

\subsubsection{Lower bound on the secure key rate}
In the noise-adaptive secure dense coding based key distribution 
protocol, when both the honest parties perform a common key generation run, the shared classical-classical-quantum ($ccq$) purified state is 
\begin{equation}\label{eq:ccq-SDC}
    \begin{aligned}
        &\kappa_{AA'BB'E} = \hat{\tilde{\mathcal{B}}}_{AA'} \otimes \hat{\mathcal{B}}_{BB'}(|\psi\rangle\langle \psi|_{AA'BB'E}) = \sum_{i,j,x,y} q(ijxy) |ij\rangle\langle ij|_{AA'} \otimes |xy\rangle\langle xy|_{BB'} \otimes \rho_E^{ijxy} \, .
    \end{aligned}
\end{equation}
where $|\psi\rangle_{AA'BB'E}$ is the pure quantum state shared by all three parties. 

Here, $\hat{\mathcal{B}}(\cdot)$ has been given in Eq. \eqref{eq:Bell-super}, and (tilde) $\hat{\tilde{\mathcal{B}}}(\rho) = |\chi(ij)\rangle \langle \chi(ij)| \rho |\chi(ij)\rangle \langle \chi(ij)|$ for all $i,j \in \{0,1\}$, where $|\chi(ij)\rangle$, given in \eqref{eq:BellAlice}, can be written as $|\chi(ij)\rangle = (I \otimes W) |B(ij)\rangle$, where $|B(ij)\rangle$ are the four maximally entangled Bell state. 
The set of $\{|\chi(ij)\rangle\}$ corresponds to the rotated Bell measurement performed by Alice in the noise-adaptive setting. After completion of the protocol, Eve's state reduces to $\rho_E^{ijxy}$ with the corresponding probability $q(ijxy)$, depending on the measurement statistics of Alice and Bob. Hence, Eve can extract information about the generated key by measuring this state. The secret key rate, $r$, even in the presence of Eve, is lower bounded by~\cite{Devetak_PRSA_2005_key-rate}
\begin{eqnarray}
    r &\geq & I(A:B)_{\kappa} - I(B:E)_{\kappa} \geq \log_2 \frac{1}{\gamma} - S(B|A)_\tau - S(B|A)_{\kappa}\label{eq:devetak-sdc}
\end{eqnarray}
where $\gamma = \max_{(ij),(xy)} || \sqrt{\mathcal{G}(ij)}\sqrt{\mathcal{B}(xy)}||_\infty^{\,2}$, and  the state $\tau$ is given by
\begin{align}
        &\tau_{AA'BB'E} = \hat{\tilde{\mathcal{G}}}_{AA'}\otimes \hat{\mathcal{G}}_{BB'} (|\psi\rangle\langle \psi|_{AA'BB'E}), \label{eq:App-tau-SDC}
\end{align}
which is another $ccq$ state shared between the honest parties and the eavesdropper, when both the honest parties perform a common test run. Note that in writing Eq.~\eqref{eq:devetak-sdc}, we have used the similar derivation steps as in Eq.~\eqref{eq:devetak-key} to Eq.~\eqref{eq:working-key-rate}, with the help of a fictitious state similar to $\xi_{AA'BB'E}$ defined in Eq.~\eqref{eq xi}.

Here, $\hat{\mathcal{G}}(ij)$ implies the test measurement performed by Bob, which takes the form $\hat{\mathcal{G}}(ij) (\rho) = |i,j_\vdash \rangle \langle i,j_\vdash|\,\rho\,|i,j_\vdash \rangle \langle i,j_\vdash|$ for all $i,j\in\{0,1\}$. The bases $|i,j_\vdash \rangle$ are the product bases and represent the eigenbasis of $\hat{\sigma}^Z \otimes \hat{\sigma}^X$, specified as
\begin{equation}
    |j,k_\vdash\rangle_{AA'} = \frac{1}{\sqrt{2}} \sum_{m=0}^{1} e^{\iota \pi km} |j,m\rangle_{AA'}\,
\end{equation}
whereas Alice's test measurement is represented by (tilde) $\hat{\tilde{\mathcal{G}}}_{AA'}(\cdot)$, whose measurement operators are given by $(\mathbb{I}\otimes W)|j,k_\vdash\rangle \langle j,k_\vdash| (\mathbb{I}\otimes W^{\dagger})$.\\

One can easily check that for this particular choice of the measurement operators performed by Bob, for both the key generation run and test runs, gives us an optimum lower bound on the secret key rate. As given in Eq. \eqref{eq:devetak-sdc}, the lower bound comprises three quantities, one is $\gamma$, which can now be expressed as 
$\gamma=\max_{j,k,x,y} |\langle jk_\vdash| B(xy)\rangle|^2$. To estimate $\gamma$, let us evaluate $\langle jk_\vdash| B(xy)\rangle$:
\begin{equation}
    \begin{aligned}\label{eq:}
        \langle jk_\vdash| B(xy)\rangle &= \frac{1}{2} \sum_{l,m=0}^{1} e^{-\iota \pi(km-ly)} \langle j,m|l,l\oplus x\rangle \\
        &= \frac{1}{2} \sum_{l,m=0}^{1} e^{-\iota \pi(km-ly)} \delta_{j,l} \delta_{m,l\oplus x}\\
        &= \frac{1}{2} e^{-\iota \pi\, (k(j\oplus x)-jy)}\,,
    \end{aligned}
\end{equation}
hence $\gamma=\max_{j,k,x,y} |\langle jk_\vdash| B(xy)\rangle|^2 = 4$, and the first quantity give us $\log_2 \frac{1}{\gamma} = 2$, the other two quantities the conditional shannon entropies should be minimized over the choice of adaptive unitary $W(\theta,\chi,\phi)$,  to get the optimal key rate, hence the final key rate is
\begin{eqnarray}
   \hspace{-0.5em}    \nonumber r_{adaptive} 
        &\geq& \max\bigg[0,\,\max_{\theta,\chi,\phi} \left( 2 - S(B|A)_\tau - S(B|A)_{\kappa}\right)\bigg].\\
        && \label{eq:SDC-key-rate-maximized}
\end{eqnarray}


\section{Entanglement-free noise adaptive two-way quantum key distribution protocol}\label{app:LM05}

In this section, we present an entanglement-free, noise-adaptive two-way QKD protocol involving qubit systems, based on a modified LM$05$ protocol~\cite{Lu_PRA_2011_LM05-secure,Beaudry_PRA_2013_two-way-QKD,Muhuri_PLA_2026}. We then outline the procedure for computing the secret key rate generated by this protocol in the presence of an eavesdropper capable of performing collective attacks.

\subsubsection{Description of the protocol}
Let us now describe the protocol which involves two honest parties, say Alice ($A$), and Bob ($B$); and an eavesdropper, say Eve ($E$). It proceeds in three main steps - the preparation stage, the encoding stage, and the measurement stage. 

{\bf Preparation:} $A$ randomly prepares a state, $\rho_A$, either from the computational basis set, $\{\ket{0},\ket{1}\}$, or the Fourier basis, $\{\ket{\tilde{0}} = \ket{+},\ket{\tilde{1}} = \ket{-}\}$. Subsequently, $A$ stores $i$ as her input bit whenever she prepares one of the states from the set $\{\ket{i},\ket{\tilde{i}}\}$. Moreover, $A$ also remembers her choice of basis in which the state is prepared.

{\bf Encoding:} Upon receiving the state $\ket{i(\tilde{i})}_A$ through a quantum channel, $\Lambda^f_{A \to B}$ (which henceforth we refer to as the {\it forward channel}), $B$ either encodes the message with a high probability, $c \approx 1$, or performs a security check with probability $(1-c)$. 
\begin{itemize}
    \item {\it Key generation run with probability $c$.} In the key generation run, $B$ performs one of the four unitary operations $\{U^{xy}= W(\theta,\chi,\phi) \hat{\sigma}^{xy}\}_{x,y\in \{0,1\}}$ on his received system for encoding, i.e., $\rho_B'^{xy}=U^{xy}\,\Lambda^f_{A \to B}(\rho_A)\,U^{xy^\dagger}$, where $\{\hat{\sigma}^{xy}\}_{x,y\in \{0,1\}}$ and $W(\theta,\chi,\phi)$ are described in the previous section. $B$, then, stores both the bits $x$ and $y$. Henceforth, we will use the shorthand notation $W$, interchangeably in place of $W(\theta,\chi,\phi)$ for brevity.
    
    \item {\it Test run with probability $1 - c$.} $B$ measures either in the computational basis, $\left(\{\ket{0},\ket{1}\}\right)$, or in the Fourier basis, i.e.,  $\{\ket{\tilde 0},\ket{\tilde 1}\}$. Following the measurement, the resulting system with post-measurement state undergoes a unitary rotation $W$ and is then sent back to $A$ through the {\it backward channel}, $\Lambda^b_{B \to A}$. 
\end{itemize}
{\bf Measurement:} The final step involves $A$ measuring $\rho''^{xy}_A = \Lambda^b_{B \to A} (\rho'^{xy}_B)$ in the case of key generation run, or, in the test run measuring the system returned by $B$ after its measurement and subsequent rotation by $W$. In both cases, the measurement is performed in the same basis, albeit with a rotation by $W$, in which $\rho_A$ was originally prepared. The resulting outcome, denoted by $o$, is then recorded as $A$'s output bit.\\

\noindent \textit{Post-processing}. After repeating the above steps $N$ times, the parties perform classical post-processing wherein $A$ reveals her choice of basis in which the initial states were prepared, while $B$ discloses which qubits were encoded and which were measured along with the measurement basis. In our work, we concentrate on reverse basis-reconciliation~\cite{Beaudry_PRA_2013_two-way-QKD}. In the key generation run, $A$ computes the key value by performing bit-wise XOR operation between $i$ and $o$, while $B$ keeps the first recorded bit, $x$, if the initial state was prepared in the computational basis or the second bit, $y$, if it was prepared in the Fourier basis. In the test runs, $A$ and $B$ discard their outcomes whenever their measurement bases do not coincide; otherwise, they can detect the presence of Eve by comparing their measurement outcomes. In particular, the absence of a perfect double-correlation between the measurement bases on the forward and backward channels, during the check runs, signals the presence of $E$.

Note that the protocol is deterministic, as no key bits are discarded in the key generation runs due to basis mismatch during post-processing. 

\subsubsection{Recipe to compute the key rate}
Let us now assume that the eavesdropper, $E$, is capable of performing a collective attack in this protocol. In such a scenario, the lower bound on the secret key rate can be evaluated using the Devetak--Winter prescription \cite{Devetak_PRSA_2005_key-rate}. As discussed in the previous section, the formulation of $ccq$ states is again required to effectively apply this bound. For this purpose, we consider purified versions of both the state preparation and encoding operations. In particular, as already outlined earlier, the encoding operations can be purified by appending a Bell state to the target system and subsequently performing a Bell basis measurement jointly on the target system and one subsystem of the appended Bell pair. Furthermore, the state preparation can also be purified by employing a pre-shared Bell state, followed by a suitable measurement on one subsystem.

Under the assumption of a collective attack, the eavesdropper is taken to hold a global pure state, $\ket{\psi}_{A A' B B' E}$~\cite{Beaudry_PRA_2013_two-way-QKD}, with different subsystems distributed among the legitimate parties. As the protocol proceeds, $A$ and $B$ perform their respective operations, while $E$ may measure her subsystem to extract information about the secret message. It is worth noting that, in the purified version of the protocol, all operations performed by the parties $A$ and $B$ are realized solely through measurements, possibly preceded, when required, by the appending of a Bell state.

A key difference from the entanglement-based protocol lies in the rate of key generation per run. In entanglement-based schemes, it is possible to generate up to two bits of key per run, whereas in the present protocol, at most one bit can be established per run. This limitation naturally necessitates the use of coarse-grained measurements in both the encoding and decoding stages~\cite{Muhuri_PLA_2026}.

Considering, $\hat{M}(i,j) = (\mathbb{I} \otimes W)\ket{ij}\bra{ij} (\mathbb{I} \otimes W^\dagger)$ and $\hat{B}(i,j)=(\mathbb{I} \otimes W^\ast)\ket{B(ij)}\bra{B(ij)}(\mathbb{I} \otimes W^T)$ with $\{i,j\}\in\{0,1\}$, we enumerate the coarse-grained encoding and decoding measurements as well as the corresponding test measurements, as follows: 
\begin{enumerate}
    \item \textbf{Computational basis runs:} \begin{eqnarray}
       && \nonumber \begin{rcases}
           \text{encoding: }\mathbb{B}_0(x)_{BB'}=\sum_{y=0}^1 \hat{B}(xy)_{BB'}, \\
           \text{decoding: }\mathbb{M}_0(x)_{AA'}=\sum_{k=0}^1 \hat{M}(k,k\oplus x)_{AA'}
       \end{rcases} ~ \text{key run}, \\
        \label{eq:coarse_comp-encoding} \\
       && \nonumber \begin{rcases}
       \mathbb{N}_0(x)_{BB'} = (W^\ast \otimes \mathbb{I}) \ket{\tilde{x}}\bra{\tilde{x}}_{B} \otimes\mathbb{I}_{B'} (W^T \otimes \mathbb{I}), \\
       \mathbb{N}_0(x)_{AA'} = (\mathbb{I} \otimes W) \mathbb{I}_A\otimes \ket{\tilde{x}}\bra{\tilde{x}}_{A'} (\mathbb{I} \otimes W^{\dagger})
       \end{rcases} ~\text{test run} \\
       \label{eq:coarse_comp-testB}
    \end{eqnarray}
    \item \textbf{Fourier basis runs:} 
    \begin{eqnarray}
       && \nonumber \begin{rcases}
           \text{encoding: }\mathbb{B}_1(y)_{BB'}=\sum_{x=0}^1 \hat{B}(xy)_{BB'}, \\
           \text{decoding: }\mathbb{M}_1(y)_{AA'}=\sum_{k=0}^1 \hat{M}(\tilde{k},\widetilde{k\oplus y})_{AA'}
       \end{rcases} ~ \text{key run}, \\
        \label{eq:coarse_fourier-encoding} \\
       && \nonumber \begin{rcases}
       \mathbb{N}_1(x)_{BB'} = (W^\ast \otimes \mathbb{I}) \ket{x}\bra{x}_{B}\otimes\mathbb{I}_{B'} (W^T \otimes \mathbb{I}),\\
       \mathbb{N}_1(x)_{AA'} = (\mathbb{I} \otimes W) \mathbb{I}_A\otimes \ket{x}\bra{x}_{A'} (\mathbb{I} \otimes W^{\dagger}) 
       \end{rcases} ~\text{test run} \\
       \label{eq:coarse_fourier-test}
    \end{eqnarray}
\end{enumerate}
As there are two different preparation bases at the disposal for $A$, denoted by $\theta = 0,1$, corresponding to the computational and Fourier bases respectively, we obtain the final state by averaging over the preparations which occur with a probability, $p_\theta$, as

\begin{eqnarray}
   \nonumber  \kappa_{A A' B B' E}  &=& \sum_{\theta=0}^1\sum_{x,x' =0}^{1} p_{\theta} \Big[  \mathbb{M}_\theta(x)_{AA'}\otimes\mathbb{B}_\theta(x')_{BB'}\Big] \Big(\rho^{\psi}_{AA'BB'E}\Big)\\&=&\sum_{\theta=0}^1\sum_{x,x' =0}^{1} p_{\theta}q_{xx'}^{\theta} \ket{x}\bra{x}_{A}\otimes \ket{x'}\bra{x'}_{B}\otimes\rho_{E}^{xx'\theta} = \sum_{\theta=0}^1 p_{\theta} \kappa^{\theta}_{AA'BB'E},
     ~~~ \label{eq:kappa_state}
\end{eqnarray}
where
$$\kappa^{\theta}_{AA'BB'E}=\sum_{x,x' =0}^{1} [  \mathbb{M}_\theta(x)_{AA'}\otimes\mathbb{B}_\theta(x')_{BB'}\Big] \Big(\rho^{\psi}_{AA'BB'E}\Big).$$ 

Note that $\ket{x}_{A}$ and $\ket{x'}_{B}$ are classical registers recording the raw keys, and $\rho_{E}^{xx'\theta}$ is the state at $E$ depending on the measurement statistics, $q^{\theta}_{x x'}$, enabling the eavesdropper to gather information about the protocol.

To derive the key rate, let us define two fictitious states 
\begin{eqnarray}
  \nonumber && \tau_{A A' B B' E} = \sum_{\theta=0}^1\sum_{x,x' =0}^{1} p_{\theta}\Big[  \mathbb{N}_\theta(x)_{AA'}\otimes\mathbb{N}_\theta(x')_{BB'}\Big] \Big(\rho^{\psi}_{AA'BB'E}\Big), =\sum_{\theta=0}^1 p_{\theta}\tau_{A A' B B' E}^{\theta}, ~
  \label{eq:tau_state} \\ \nonumber \\
   \nonumber ~\text{and}~~ && \xi_{A A' B B' E} =\sum_{\theta=0}^1\sum_{x,x' =0}^{1} p_{\theta}\Big[ \mathbb{N}_\theta(x)_{AA'}\otimes \mathbb{B}_\theta(x')_{BB'} \Big] \Big(\rho^{\psi}_{AA'BB'E}\Big)= \sum_{\theta=0}^1 p_{\theta}\xi_{A A' B B' E}^{\theta}.
   \label{eq:sigma_state}
\end{eqnarray}

The lower bound on the key rate, $r$, is an average rate over two preparation bases, i.e., $r = \sum_{\theta=0}^1 p_{\theta}r^{\theta}$, where $r^{\theta}$ represents the lower bound on the key rate corresponding to preparation in the $\theta$ basis  
\begin{eqnarray}
   r^{\theta} &\ge& I(A:B)^{\Theta=\theta}_{\kappa} - I(B:E)^{\Theta=\theta}_{\kappa} \label{eq:key-rate_1}\\
    &=& S(B|E)^{\Theta=\theta}_{\kappa} - S(B|A)^{\Theta=\theta}_{\kappa} \label{eq:key-rate_2} \\
    &=& \textcolor{black}{S(B|E)^{\Theta=\theta}_{\xi} - S(B|A)^{\Theta=\theta}_{\kappa} \label{eq:key-rate_3}} \\
    &\geq& \log_2 \frac{1}{\gamma} - S(B|A)^{\Theta=\theta}_{\tau} - S(B|A)^{\Theta=\theta}_{\kappa} \label{eq:key-rate_4}.
\end{eqnarray}
In this protocol, $p_\theta$ is assumed to be $\frac 1 2$ for both preparations,
and the entropic uncertainty relation\footnote{Corresponding to two different POVM settings, denoted as $P_X\equiv\{P^i_X\}_i$ and $P_Z \equiv\{P^j_Z\}_j$, with classical outcomes, $i$ and $j$ respectively, performed by the party $B$ on a tripartite state $\rho_{ABE}$, the entropic uncertainty relation states as
\begin{equation}
    S(X|E)+S(Z|A)\geq \log_2 \frac{1}{\gamma},
\end{equation}
where $\gamma = \max_{i,j} ||\sqrt{P^i_X}\sqrt{P^j_Z}||_\infty^2$~\cite{ Berta_NP_2010_uncertainty-principle-memory, Coles_RMP_2017_entropic-uncertainty-relations-review}.}, used here is $S(B|E)_{\xi} + S(B|A)_{\tau} \geq \log_2 \frac 1 \gamma$ with $\gamma = \max_{x, x'} ||\sqrt{\hat{\mathcal{B}_\theta}(x)} \sqrt{\hat{\mathcal{N}_\theta}(x')}||_{\infty}^2 $, to derive the final expression for the lower bound on the key rate. \\
One can easily check that here $\gamma = 2$, hence the final key rate for \emph{noise adaptive} LM05 protocol is 
\begin{eqnarray}
   \hspace{-0.5em}    \nonumber && r_{adaptive} \geq
          \max\bigg[0,\,\max_{\theta,\chi,\phi} \left( 1 - \frac 12\sum_{\theta=0}^1 \big(S(B|A)^{\Theta=\theta}_{\tau} - S(B|A)^{\Theta=\theta}_{\kappa}\big)\right)\bigg].\\
        && \label{eq:LM05-key-rate-maximized}
\end{eqnarray}

\section{BB84 Quantum Key Distribution Protocol}\label{sec:secureBB84}

The BB84 protocol \cite{BB84} enables two distant parties, Alice and Bob, to establish a shared secret key over an insecure quantum channel supplemented by an authenticated classical channel.

\textbf{State Preparation:}
Alice prepares a sequence of quantum states chosen randomly from the set 
\begin{equation}
\left\{ \{\ket{0}, \ket{1}\}, \{\ket{+}, \ket{-}\} \right\},
\end{equation}
where $\ket{0}, \ket{1}$ are the eigenbasis of $\hat{\sigma}^Z$, and 
\begin{equation}
\ket{\pm} = \frac{1}{\sqrt{2}}\left(\ket{0} \pm \ket{1}\right),
\end{equation}
are the eigenbasis of $\hat{\sigma}^X$. Alice stores the bit value $i = 0$, if she prepares $\ket{0}$ or $\ket{+}$ and $i = 1$, for $\ket{1}$ or $\ket{-}$. 
After preparation of the signal qubit, 
Alice transmits the resulting quantum states to Bob through a quantum channel $\Lambda$. Upon receiving each signal, Bob independently chooses a random measurement basis $M_i \in \{\hat{\sigma}^Z,\hat{\sigma}^X\}$ and performs a projective measurement, obtaining an outcome $x \in \{0,1\}$.

\textbf{Sifting Procedure:} After repeating the preparation and measurement procedure $N$ times, the honest parties perform the basis reconciliation or the shifting.  
In this part, Alice and Bob publicly announce their respective basis choices, either $\hat{\sigma}^Z$ or $\hat{\sigma}^X$, over an authenticated classical channel. They retain only those runs of the protocol for which their basis choices match, and discard all the others. The remaining data define the \emph{sifted key}, and will be further used for one-way classical post-processing runs for distilling the final raw key.

\subsection{Computing the key rate}
One can easily check that Alice's preparation of the signal qubit can be equivalently considered as a measurement procedure on a shared maximally entangled Bell state. 
An equivalent formulation of the BB84 protocol can be given in an entanglement-based picture, which is particularly useful for security analysis. This version is closely related to the protocol introduced by \cite{Ekert_PRL_1991_Ekert-protocol}.
Here, Alice prepares a  maximally entangled Bell state, $\ket{\phi^+}_{AA'}$, keeps one part $A$, in her quantum memory and sends the other ($A'$) to Bob via a quantum channel $\Lambda^f_{A' \rightarrow B}$, resulting in the shared state $\rho_{AB} =\Lambda^f_{A' \rightarrow B} (\rho_{AA'})$, with $\rho_{AA'} = |\phi^+\rangle \langle\phi^+|_{AA'}$. \\[2pt]

\textbf{Measurements by both parties:}
For each shared pair, $\rho_{AB}$, Alice and Bob independently and randomly choose a measurement basis from the set $\{\hat{\sigma}^Z, \hat{\sigma}^X\}$, and perform
projective measurements on their respective qubits, obtaining outcomes $i, x \in \{0,1\}$.

Note that Alice's measurement in $\hat{\sigma}^Z (\hat{\sigma}^X)$, basis and obtaining the outcome $i$, is equivalent to preparing the qubit in the respective eigenbasis $\ket{i} (\ket{i_\vdash})$, and sending it through $\Lambda^f_{A' \rightarrow B}$. 
Without any loss of generality, we can set beforehand that the measurements by the two parties in the $\hat{\sigma}^Z$ basis correspond to the \emph{key generation run} and measurement in the $\hat{\sigma}^X$ basis to the \emph{test run}. In the case of the noise-adaptive protocol, the changes will occur to the measurement schemes of both parties, which in turn implies the preparation of a rotated resource state and its measurement in a rotated basis. Alice will prepare states like $W\ket{i}$ or $W\ket{\tilde{i}}$, which will pass through the noisy channel and be measured by Bob in the rotated $\hat{\sigma}^Z$ or $\hat{\sigma}^X$ basis. Hence, after Bob's measurement in the rotated $\hat{\sigma}^Z$ basis (corresponding to the \emph{key run}), he creates a classical-classical ($cc$) state of the form
\begin{eqnarray}\label{eq: bb84 adap cc1}
    \nonumber && \kappa_{A B E}\\
    \nonumber &=& \sum_{i,x=0}^{1} \hat{\mathcal{P}}(i)_A \otimes \hat{\mathcal{M}}(x)_B \Big(\rho^\psi_{ABE}\Big)\\
    &=& \sum_{i,x=0}^{1} p(ix) \,\ket{i}\bra{i}_A \otimes \ket{x}\bra{x}_B \otimes \rho_{E}^{ix},
\end{eqnarray}
where we have considered the rotated measurement bases for the adaptive protocol as
\begin{eqnarray}\label{eq:bb84 adaptive key}
    \hat{\mathcal{P}}(i) &=& W^{\ast} \ket{i}\bra{i}W^T,\\
    \hat{\mathcal{M}}(i) &=& W \ket{i}\bra{i} W^{\dagger}.
\end{eqnarray}
In Eq.~\eqref{eq: bb84 adap cc1}, $\rho_{ABE}^{\psi}$ denotes the density matrix corresponding to $\ket{\psi}_{ABE}$; a global pure state held by the eavesdropper under the assumption of a collective attack.

Note that $\ket{i}_A$ and $\ket{x}_B$ represent the classical registers containing Alice's and Bob's measurement outcomes with the corresponding joint probability distribution denoted by $p(ix) = p(i)p(x|i)$. 

Similarly, we can consider the $cc$ state resulting from the test run (rotated $\hat{\sigma}^X$ measurements) given by
\begin{equation}
    \tau_{A B E} = \sum_{i,x =0}^{1} \hat{\mathcal{P}}^f(i)_A \otimes \hat{\mathcal{M}}^f(x)_B \Big(\rho^\psi_{ABE}\Big),
\end{equation}
where the test measurement bases are
\begin{eqnarray}\label{eq:bb84 adaptive test}
    \hat{\mathcal{P}}^f(i) &=& W^{\ast} \ket{\tilde{i}}\bra{\tilde{i}}W^T,\\
    \hat{\mathcal{M}}^f(i) &=& W \ket{\tilde{i}}\bra{\tilde{i}} W^{\dagger}.
\end{eqnarray}

To derive the key rate, we define another fictitious state as follows 
\begin{equation}
    \xi_{A B E} = \sum_{i,\,x =0}^{1} \hat{\mathcal{P}}^f(i)_A \otimes \hat{\mathcal{M}}(x)_B \Big(\rho^\psi_{ABE}\Big).
\end{equation}
The lower bound on the secret key rate, $r$, can be derived similarly as done in earlier section \ref{app:LM05}, following~\cite{Devetak_PRSA_2005_key-rate}
\begin{eqnarray}
    r &\geq& I(A:B)_\kappa
    - I(B:E)_\kappa\label{eq:bb84 key-rate_1}\\
    &=& S(B|E)_{\kappa} - S(B|A)_{\kappa}\label{eq:bb84 key-rate_2}\\
    &=& S (B|E)_\xi - S(B|A)_{\kappa}\label{eq:bb84 key-rate_3}\\
    &\geq& \log_2
    \frac{1}{\gamma}-S(B|A)_\tau - S(B|A)_{\kappa}.\label{eq:bb84 key-rate_4}
\end{eqnarray}
In Eqs. \eqref{eq:bb84 key-rate_1}-\eqref{eq:bb84 key-rate_4}, the superscripts denote the states for which the quantities like mutual information ($I$) and von Neumann entropy terms ($S$) are calculated. The definitions of the entropy terms are the same as given earlier in section~\ref{sec:generic-key-rate}, below Eq.~\eqref{eq:devetak-key}. In deriving Eq.~\eqref{eq:bb84 key-rate_3}, we have used the fact that $S(B|E)_\kappa=S(B|E)_\xi$, evident from the definitions of the states $\kappa_{A B E}$ and $\xi_{A B E}$, which differ in the measurement performed by Alice, that does not affect the conditional entropy between $B$ and $E$. We have also used the entropic uncertainty relation $S(B|E)_\xi+S(B|A)_\tau \geq \log_2 \frac{1}{\gamma}$, where $\gamma=\max_{i,x} ||\sqrt{\hat{\mathcal{M}}(i)} \sqrt{\hat{\mathcal{M}}^f(x)} ||_{\infty}^2=\frac{1}{2}$.\\
It implies that for fully correlated measurements performed by Alice and Bob to obtain $\kappa_{A B E}$ and $\tau_{A B E}$, the two von Neumann entropy terms in \eqref{eq:bb84 key-rate_4} vanish. Hence, the lower bound for the secure key rate reduces to $r\geq \log_2 (1/ \gamma) = 1$.

We have considered a two-way variant of the BB84 protocol, solely for the purpose of comparison with the key rates of the two fundamentally deterministic two-way protocols, i.e. the SDC and the LM05 protocols. This can be thought of as a merger of two separate BB84 protocols (standard one-way types), in which we have chosen the legitimate party preparing the resource qubit to differ in the different runs. As a result, the noisy qubit, generated after traveling through the noisy quantum channel, 
is now measured by different parties. For Pauli channels, the key rate expressions for the two runs would be identical; therefore, the final key rate for the two-way variant is twice the one-way key rate. However, for the Amplitude-damping channel, this is not the case, as this channel acts differently on qubit states $|0\rangle$ and $|1\rangle$. Here, the final key rate for the two-way version needs to be computed by adding the two rates originating from the separate one-way runs.\\

\end{widetext}

\appendix

\end{document}